\documentclass[twocolumn,prb,showpacs,floatfix]{revtex4}

\usepackage{graphicx}
\usepackage{dcolumn}
\usepackage{bm}
\usepackage{amsmath}

\begin{document}

\title{Artificial quantum-dot Helium molecules: \\ Electronic spectra, spin
structures, and Heisenberg clusters}

\author{Ying Li}
\author{Constantine Yannouleas}
\email{Constantine.Yannouleas@physics.gatech.edu}
\author{Uzi Landman}
\email{Uzi.Landman@physics.gatech.edu}

\affiliation{School of Physics, Georgia Institute of Technology,
             Atlanta, Georgia 30332-0430}

\date{6 May 2009; Physical Review B, {\bf in press}}

\begin{abstract}

Energy spectra and spin configurations of a system of $N=4$ electrons in lateral 
double quantum dots (quantum dot Helium molecules) are investigated 
using exact diagonalization (EXD), as a function of interdot separation, applied 
magnetic field $(B)$, and strength of interelectron repulsion. 
As a function of the magnetic field, the energy spectra exhibit a low-energy 
band consisting of a group of six states, with the number six being a 
consequence of the conservation of the total spin and the ensuing spin 
degeneracies for four electrons. The energies of the six states appear to cross 
at a single value of the magnetic field, and with increasing Coulomb 
repulsion they tend to become degenerate, with a well defined energy gap 
separating them from the higher-in-energy excited states. The appearance of the 
low-energy band is a consequence of the formation of a Wigner supermolecule, 
with the four electrons (two in each dot) being localized at the vertices of a 
rectangle. Using spin-resolved pair-correlation distributions, a method for 
mapping the complicated EXD many-body wave functions onto simpler spin functions 
associated with a system of four localized spins is introduced. Detailed 
interpretation of the EXD spin functions and EXD spectra associated with the 
low-energy band via a 4-site Heisenberg cluster (with $B$-dependent 
exchange integrals) is demonstrated. Aspects of spin entanglement, referring to 
the well known $N$-qubit Dicke states, are also discussed.
\end{abstract}


\maketitle

\section{Introduction}  

The field of two-dimensional (2D) semiconductor quantum dots (QDs) has witnessed
rapid expansion in the last several years, both experimentally 
\cite{kouw01,hans07} and theoretically.\cite{maks00,reim02,yann07} Along with
fundamental interest in the properties of such systems, and as a test ground
for highly correlated electrons, a major motivation for these growing endeavors 
has been the promising outlook and potential of quantum dots concerning the 
implementation of solid-state quantum computing and quantum information devices.
\cite{loss98,loss99,zuti04,awsc07,lida06} To this effect highly precise control 
of the space and spin degrees of freedom of 
a small number $N$ of confined electrons 
(down to an empty \cite{kouw96,cior00,heib05} QD) needs to be achieved, and 
experimentally this was demonstrated recently for two electrons in a lateral 
double quantum dot molecule (see Ref.\ \onlinecite{hans07}, and references
therein). From the theoretical standpoint, high-level computational methods that
reach beyond the level of mean-field approximation are needed, \cite{yann07} 
with the ability to provide solutions that preserve all the symmetries 
of the many-body Hamiltonian, and in particular those associated with the
total spin. In this context, electrons in quantum dots exhibit localization in 
space and formation of Wigner molecules (see, e.g., Ref.\ \onlinecite{yann07}).
When the spin degree of freedom is considered, such Wigner molecules
may be viewed as finite Heisenberg spin clusters 
\cite{hend93,haas07} whose quantum behavior (due to finite-size fluctuations and
correlation effects) differs drastically from the behavior expected from 
magnetic systems in the thermodynamic limit.\cite{hend93,fazebook} 

There is an abundance of experimental and theoretical publications concerning 
{\it circular single\/} quantum dots with a small number of electrons. 
\cite{maks00,reim02,kouw01,yann07,peet03,ront06,umri07} In this paper, we use 
exact diagonalization \cite{reim02,yann07,ront06} (EXD) to investigate the 
properties of {\it lateral double quantum dots\/} (DQDs) containing four 
electrons. DQDs are referred to also as artificial molecules. Specifically in 
the case of four electrons they can be viewed as artificial {\it quantum dot
Helium molecules\/}.\cite{note8} 
DQDs containing two electrons have been already studied extensively both 
experimentally \cite{hans07} and theoretically.\cite{yann07,hell05,lebu06} 
However, experimental studies of DQDs with more than two electrons are 
relatively few.\cite{chan04,taru08} We are aware of a single theoretical EXD 
study of a lateral DQD with three electrons,\cite{peet05} and another one of two
laterally coupled quantum rings with three electrons.\cite{szaf08}

In light of the novel quantum behavior discovered in our investigations (compared
to circular QDs with regard to the spectra, spin structures, and analogies with
Heisenberg clusters), we hope that the present work would serve as an impetus 
for further experimental studies on lateral DQDs. In particular, as a function 
of the magnetic field, we find that: (1) A low-energy band of six states 
develops as the strength of the Coulomb repulsion increases, separated by an 
energy gap from the other excited states, and (2) All six states appear to 
``cross'' at a single value of the magnetic field. The crossing point gets 
sharper for larger interdot distances. We find that the specific number of 
crossing states (six) derives from the spin degeneracies and multiplicities 
expressed by the branching diagram.\cite{pauncz}

The formation of the low-energy band is a consequence of the localization of
the four electrons within each dot (with two electrons in each dot). This
localization leads to formation (with increasing strength of the Coulomb 
repulsion) of a Wigner supermolecule,\cite{yann99} with the four localized 
electrons (two in each dot) being located at the corners of a rectangular 
parallelogram (RP). Using the spin-resolved pair-correlation
functions, we show how to map the EXD many-body wave functions onto the spin 
functions associated with the four localized spins. This mapping leads us
naturally to study analogies with finite systems described by a model 
Heisenberg Hamiltonian (often referred to as finite Heisenberg clusters).
Specifically, we provide a detailed interpretation of the EXD spin functions and
EXD spectra associated with the low-energy band via a 4-site finite Heisenberg 
cluster characterized by two (intradot and interdot) exchange integrals. 

More importantly, our EXD calculations exhibit a prominent oscillatory 
magnetic-field dependence of the two exchange integrals entering in this 4-site 
Heisenberg Hamiltonian. Such strong $B$-dependence 
of the exchange integrals has been found
in previous theoretical studies in the simpler case of two electrons in double 
quantum dots, \cite{loss99, hell05,lebu06,yann02,yann02.2} as well as in 
anisotropic single quantum dots.\cite{yann06,yann07.2} 
The $B$-dependence in the case of two electrons in quantum dots has also been
observed experimentally.\cite{yann06,yann07.2,marc04} Following an earlier 
proposal,\cite{loss99} the $B$-dependence of the exchange integral for the 
two-electron case has developed into a central theme in experimental efforts 
aiming at solid-state implementation of quantum computing.\cite{hans07,taru08}
Our EXD results in this paper extend the $B$-dependence of the exchange 
integrals to larger numbers $(N > 2)$ of electrons in quantum dot molecules.

We further discuss that the determination of the equivalent spin functions 
enables consideration of aspects of entanglement regarding the EXD 
solutions. In particular, we show that the formation of Wigner supermolecules
leads to strongly entangled states known in the literature of quantum 
information as $N$-qubit Dicke states.\cite{dick54,vers02,stoc03,korb06}

We finally mention that the trends in the excitation spectra (e.g., formation
of a low-energy band) and entanglement characteristics (e.g., mapping to spin 
functions of localized electrons) found in the case of double quantum dots 
have many analogies with those found in other deformed configurations, and in 
particular single anisotropic quantum dots; see, e.g., the case of three 
electrons in Ref.\ \onlinecite{yues07}.

The plan of the paper is as follows:
\begin{itemize}
\item
Section II describes the two-dimensional two-center-oscillator (TCO) external 
confining potential that models the double quantum dot. 
\item
Section III reviews the many-body Hamiltonian and the exact-diagonalization 
method as implemented in this paper. 
\item
Section IV outlines some theoretical background regarding the general form of 
four-electron spin eigenfunctions and the branching diagram 
(which describes the break-down of spin multiplicities for given $N$).
\item
Section V describes our numerical results from the exact diagonalization,
that is, the EXD spectra, the electron densities, and the spin-resolved
conditional probability distributions (CPDs). 
\item
Section VI provides an interpretation of the numerical EXD results for the 
6-state lower-energy band with the help of a 4-site Heisenberg Hamiltonian. 
\item
Section VII contains two parts which discuss (a) the importance of 
$B$-dependent exchange integrals in the Heisenberg Hamiltonian and (b) the
aspects of entanglement exhibited by the EXD wave functions. 
\item 
Section VIII offers a summary.
\item 
Finally, the Appendix describes the single-particle spectrum of the two-center 
oscillator (when the two-body Coulomb interaction is omitted; non-interacting 
model). 
\end{itemize}

%
\begin{figure}[t]
\centering{
\includegraphics[width=6.5cm]{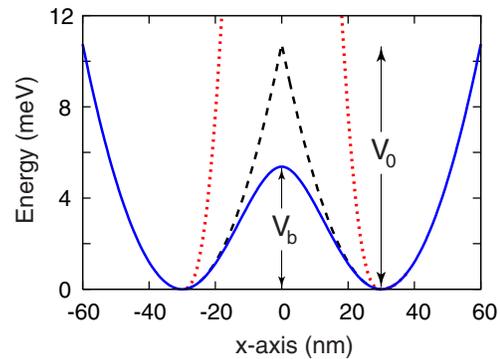}}
\caption{(Color online) Cuts at $y=0$ of the two-center-oscillator confining 
potential along the $x$-axis for an interdot separation of $d=60$ nm. 
Solid line (blue): double-oscillator confining potential with a smooth 
connecting neck specified by $\epsilon^b=V_b/V_0=0.5$. 
Dashed line (black): double-oscillator confining potential
without a smooth connecting neck. Dotted line (red): double-oscillator confining 
potential with a smooth connecting neck specified by $\epsilon^b=6$. The 
remaining parameters are 
$\hbar \omega_{x1}=\hbar \omega_{x2}=\hbar \omega_0=5.1$ meV, 
$m^*=0.070m_e$, $h_1=h_2=0$, $V_{01}=V_{02}=V_0$, 
$\epsilon^b_1=\epsilon^b_2=\epsilon^b$.
}
\label{tco_pot}
\end{figure}

\section{Two-dimensional two-center-oscillator confining potential}  
\label{2dhopot}

In the two-dimensional two-center-oscillator, the single-particle levels 
associated with the confining potential of the artificial molecule are 
determined by the single-particle Hamiltonian \cite{note1}
\begin{eqnarray}
H=T &+& \frac{1}{2} m^* \omega^2_y y^2
    + \frac{1}{2} m^* \omega^2_{x k} x^{\prime 2}_k \nonumber \\
    &+& V_{neck}(x) +h_k+ \frac{g^* \mu_B}{\hbar} {\bf B \cdot
    s}~,
\label{hsp}
\end{eqnarray}
where $x_k^\prime=x-x_k$ with $k=1$ for $x<0$ (left) and $k=2$
for $x>0$ (right), and the $h_k$'s control the relative well-depth, thus
allowing studies of hetero-QDMs. $y$ denotes the coordinate perpendicular to the
interdot axis ($x$). $T=({\bf p}-e{\bf A}/c)^2/2m^*$, with 
${\bf   A}=0.5(-By,Bx,0)$, and the last term in Eq. (\ref{hsp}) is the Zeeman 
interaction with $g^*$ being the effective $g$ factor, $\mu_B$ the Bohr 
magneton, and ${\bf s}$ the spin of an individual electron.
The most general shapes described by $H$ are two semiellipses connected by a 
smooth neck $V_{neck}(x)$ (see solid line in Fig.\ \ref{tco_pot}). $x_1 <
0$ and $x_2 > 0$ are the centers of these semiellipses, $d=x_2-x_1$
is the interdot distance, and $m^*$ is the effective electron mass.

For the smooth connecting neck, we use 
$V_{neck}(x) = \frac{1}{2} m^* \omega^2_{x k} 
[{\cal C}_k x^{\prime 3}_k + {\cal D}_k x^{\prime 4}_k] \theta(|x|-|x_k|)$, 
where $\theta(u)=0$ for $u>0$ and $\theta(u)=1$ for $u<0$.
The four constants ${\cal C}_k$ and ${\cal D}_k$ can be expressed via two
parameters, as follows: ${\cal C}_k= (2-4\epsilon_k^b)/x_k$ and
${\cal D}_k=(1-3\epsilon_k^b)/x_k^2$, where the barrier-control parameters
$\epsilon_k^b=(V_b-h_k)/V_{0k}$ are related to the actual (controlable) height 
of the bare interdot barrier ($V_b$) between the two QDs, and 
$V_{0k}=m^* \omega_{x k}^2 x_k^2/2$ (for $h_1=h_2$, $V_{01}=V_{02}=V_0$).

The single-particle levels of $H$, including an external perpendicular magnetic 
field $B$, are obtained by numerical diagonalization in a 
(variable-with-separation) basis consisting of the eigenstates of the auxiliary 
(zero-field) Hamiltonian:
\begin{equation}
H_0=\frac{{\bf p}^2}{2m^*} + \frac{1}{2} m^* 
    \omega_y^2 y^2
      + \frac{1}{2} m^* \omega_{xk}^2 x_k^{\prime 2}+h_k~.
\label{h0}
\end{equation}
The eigenvalue problem associated with the auxiliary Hamiltonian [Eq.\ 
(\ref{h0})] is separable in $x$ and $y$, i.e., the wave functions 
are written as 
\begin{equation}
\varphi_i (x,y)= X_\mu (x) Y_n (y),
\label{spphi}
\end{equation}
with $i \equiv \{\mu,n\}$, $i=1,2,\ldots,K$.

The $Y_n (y)$ are the eigenfunctions of a one-dimensional oscillator, and the
$X_\mu (x \leq 0)$ or $X_\mu (x>0)$ can be expressed through the parabolic
cylinder functions \cite{yl95,grei72} $U[\gamma_k, (-1)^k \xi_k]$, where
$\xi_k = x^\prime_k \sqrt{2m^* \omega_{xk}/\hbar}$, 
$\gamma_k=(-E_x+h_k)/(\hbar \omega_{xk})$, 
and $E_x=(\mu+0.5)\hbar \omega_{x1} + h_1$ denotes the $x$-eigenvalues.
The matching conditions at $x=0$ for the left and 
right domains yield the $x$-eigenvalues and the eigenfunctions
$X_\mu (x)$. The $n$ indices are integer. The number of $\mu$ indices is 
finite; however, they are in general real numbers.

In the Appendix, we discuss briefly the energy spectra associated with
the single-particle states of the two-center oscillator Hamiltonian
given by Eq.\ (\ref{hsp}). We follow there the notation presented first in
Ref.\ \onlinecite{yl99}. For further details, see Ref.\ \onlinecite{note2}.

In this paper, we will limit ourselves to QDMs with $x_2=-x_1$ and
$\hbar \omega_y=\hbar \omega_{x1}=\hbar \omega_{x2}=\hbar \omega_0$.
However, in several instances we will compare with the case of a single
elliptic QD where $x_2=-x_1=0$ and 
$\hbar \omega_y \neq \hbar \omega_{x} =\hbar \omega_{x1}=\hbar \omega_{x2}$.
In all cases, we will use $\hbar \omega_0 =5.1$ meV,
$m^*=0.070 m_e$ (this effective-mass value corresponds to GaAs), and
$K=50$ (which guarantees numerical convergence\cite{note5}).

\section{The Many-Body Hamiltonian and the exact diagonalization method}
\label{sec:3}
The many-body Hamiltonian ${\cal H}$ for a dimeric QDM comprising $N$ 
electrons can be expressed as a sum of the single-particle part
$H(i)$ defined in Eq.\ (\ref{hsp}) and the two-particle interelectron Coulomb
repulsion,
\begin{equation}
{\cal H}=\sum_{i=1}^{N} H(i) +
\sum_{i=1}^{N} \sum_{j>i}^{N} \frac{e^2}{\kappa r_{ij}}~,
\label{mbh}
\end{equation}
where $\kappa$ is the dielectric constant and $r_{ij}$ denotes
the relative distance between the $i$ and $j$ electrons.

As we mentioned in the introduction, we will use the method of exact
diagonalization for determining \cite{note2} the solution of the many-body 
problem specified by the Hamiltonian (\ref{mbh}). 

In the EXD method, one writes the many-body wave function 
$\Phi^{\text{EXD}}_N ({\bf r}_1, {\bf r}_2, \ldots , {\bf r}_N)$ as a linear
superposition of Slater determinants 
$\Psi^N({\bf r}_1, {\bf r}_2, \ldots , {\bf r}_N)$ that span the many-body
Hilbert space and are constructed out of the single-particle 
{\it spin-orbitals\/} 
\begin{equation}
\chi_j (x,y) = \varphi_j (x,y) \alpha, \mbox{~~~if~~~} 1 \leq j \leq K,
\label{chi1}
\end{equation}
and 
\begin{equation}
\chi_j (x,y) = \varphi_{j-K} (x,y) \beta, \mbox{~~~if~~~} K < j \leq 2K, 
\label{chi2}
\end{equation}
where $\alpha (\beta)$ denote up (down) spins. Namely
\begin{equation}
\Phi^{\text{EXD}}_{N,q} ({\bf r}_1, \ldots , {\bf r}_N) = 
\sum_I C_I^q \Psi^N_I({\bf r}_1, \ldots , {\bf r}_N),
\label{mbwf}
\end{equation}
where 
\begin{equation}
\Psi^N_I = \frac{1}{\sqrt{N!}}
\left\vert
\begin{array}{ccc}
\chi_{j_1}({\bf r}_1) & \dots & \chi_{j_N}({\bf r}_1) \\
\vdots & \ddots & \vdots \\
\chi_{j_1}({\bf r}_N) & \dots & \chi_{j_N}({\bf r}_N) \\
\end{array}
\right\vert,
\label{detexd}
\end{equation}
and the master index $I$ counts the number of arrangements 
$\{j_1,j_2,\ldots,j_N\}$ under the restriction that 
$1 \leq j_1 < j_2 <\ldots < j_N \leq 2K$. Of course, $q=1,2,\ldots$ counts
the excitation spectrum, with $q=1$ corresponding to the ground state.

The exact diagonalization of the many-body Schr\"{o}dinger equation
\begin{equation}
{\cal H} \Phi^{\text{EXD}}_{N,q}=
E^{\text{EXD}}_{N,q} \Phi^{\text{EXD}}_{N,q}
\label{mbsch}
\end{equation}
transforms into a matrix diagonalizatiom problem, which yields the coefficients 
$C_I^q$ and the EXD eigenenergies $E^{\text{EXD}}_{N,q}$. Because the
resulting matrix is sparse, we implement its numerical diagonalization 
employing the well known ARPACK solver.\cite{arpack}

The matrix elements $\langle \Psi_N^{I} | {\cal H} | \Psi_N^{J} \rangle$
between the basis determinants [see Eq.\ (\ref{detexd})] are calculated using
the Slater rules.\cite{szabbook} Naturally, an important ingredient in this
respect are the two-body matrix elements of the Coulomb interaction,
\begin{equation}
\frac{e^2}{\kappa} 
\int_{-\infty}^{\infty} \int_{-\infty}^{\infty} d{\bf r}_1 d{\bf r}_2
\varphi^*_i({\bf r}_1) \varphi^*_j({\bf r}_2)
\frac{1}{|{\bf r}_1 - {\bf r}_2|}
\varphi_k({\bf r}_1) \varphi_l({\bf r}_2),
\label{clme}
\end{equation}
in the basis formed out of the single-particle spatial orbitals 
$\varphi_i({\bf r})$, $i=1,2,\ldots,K$ [Eq.\ (\ref{spphi})]. In our approach, 
these matrix elements are determined numerically and stored separately.
Varying the dielectric constant $\kappa$ and/or the interdot-barrier parameter 
$\epsilon^b$ does not require a recalculation of the Coulomb-interaction matrix 
elements, as long as the remaining parameters are kept the same.

The Slater determinants $\Psi^N_I$ [see Eq.\ (\ref{detexd})] conserve the
third projection $S_z$,  but not the square $\hat{\bf S}^2$ of the total spin. 
However, because $\hat{\bf S}^2$ commutes with the many-body Hamiltonian, the
EXD solutions are automatically eigenstates of $\hat{\bf S}^2$ with eigenvalues
$S(S+1)$. After the diagonalization, these eigenvalues are determined by
applying $\hat{\bf S}^2$ onto $\Phi^{\text{EXD}}_{N,q}$ and using the relation
\begin{equation}
\hat{{\bf S}}^2 \Psi^N_I = 
\left [(N_\alpha - N_\beta)^2/4 + N/2 + \sum_{i<j} \varpi_{ij} \right ] 
\Psi^N_I,
\end{equation}
where the operator $\varpi_{ij}$ interchanges the spins of electrons $i$ and 
$j$ provided that their spins are different; $N_\alpha$ and $N_\beta$ denote 
the number of spin-up and spin-down electrons, respectively.

Of great help in reducing the size of the matrices to be diagonalized is the
fact that the parity (with respect to the origin) of the EXD many-body wave 
function is a good quantum number for all values of the magnetic field when
$h_1=h_2$. Specifically, the $xy$-parity operator associated with reflections 
about the origin of the axes is defined as
\begin{equation}
\hat{\cal P}_{xy} 
\Phi^{\text{EXD}}_{N,q}({\bf r}_1, {\bf r}_2, {\bf r}_3, {\bf r}_4) =
\Phi^{\text{EXD}}_{N,q} (-{\bf r}_1, -{\bf r}_2, -{\bf r}_3, -{\bf r}_4)
\label{xypar}
\end{equation}
and has eigenvalues $\pm 1$.

One can also consider partial parity operators $\hat{\cal P}_{x}$ and
$\hat{\cal P}_{y}$ associated solely with reflections about the $x$ and $y$ axis,
respectively; of course $\hat{\cal P}_{xy} = \hat{\cal P}_{x} \hat{\cal P}_{y}$.
We note that unlike $\hat{\cal P}_{xy}$, the partial parities $\hat{\cal P}_{x}$
and $\hat{\cal P}_{y}$ are conserved only for zero magnetic fields ($B=0$).
With the two-center oscillator {\it cartesian\/} basis that we use [see Eq.\ 
(\ref{spphi})], it is easy to calculate the parity eigenvalues for the Slater 
determinants, Eq.\ (\ref{detexd}), that span the many-body Hilbert space. 
Because $X_\mu(x)$ and $Y_n(y)$ conserve the partial $\hat{\cal P}_{x}$ and 
$\hat{\cal P}_{y}$ parities, respectively, one finds:
\begin{equation}
\hat{\cal P}_{xy} \Psi^N_I = (-)^{\sum_{i=1}^4 m_i+n_i} \Psi^N_I,
\label{parval}  
\end{equation}
where $m_i$ and $n_i$ count the number of single-particle states associated
with the bare two-center oscillator [see the auxiliary Hamiltonian $H_0$ in Eq.\
(\ref{h0})] along the $x$ axis and the
simple oscillator along the $y$ direction (with the assumption that the
lowest states have $m=0$ and $n=0$, since they are even states).
We note again that the index $\mu$ in Eq.\ (\ref{spphi}) is not an integer in
general, while $m$ here is indeed an integer (since it counts the number of
single-particle states along the $x$ direction).

%
\begin{figure}[t]
\centering{
\includegraphics[width=6.5cm]{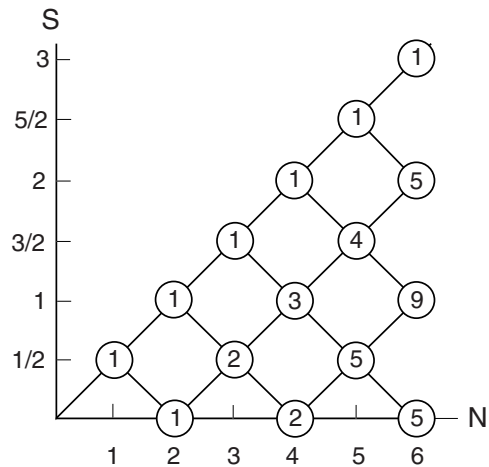}}
\caption{The branching diagram for the spin degeneracies. The total-spin
quantum number $S$ is given on the vertical axis, and the number of particles,
$N$, on the horizontal one. The numbers inside the circles give the number,
$g(N,S)$, of linear independent (and orthogonal) spin functions for the
corresponding values of $N$ and $S$.
}
\label{brandiag}
\end{figure}

\section{Many-body spin eigenfunctions}
\label{mbsfs}

For completeness and for the reader's convenience, we outline in this section
several well established (but often not well known) properties of the many-body 
spin eigenfunctions which are useful for analyzing the trends and behavior
of the spin multiplicities exhibited by the EXD wave functions for $N=4$ 
electrons. We stress here that the ability to describe spin multiplicities is
an advantage of the EXD method compared to the more familiar spin-density
functional approaches whose single-determinantal wave functions preserve only
the third projection $S_z$ of the total spin, and thus are subject to 
``spin contamination'' errors. As we will discuss below, the spin multiplicities
of the EXD wave functions lead naturally to formation of highly entangled
Dicke states,\cite{dick54,vers02,stoc03,korb06} and most importantly to 
analogies with finite Heisenberg clusters.\cite{hend93,haas07}

A basic property of spin eigenfunctions is that they exhibit degeneracies 
for $N>2$, i.e., there may be more than one linearly independent (and 
orthogonal) spin functions that are simultaneous eigenstates of both 
$\hat{\bf S}^2$ and $S_z$. These degeneracies are usually visualized by 
means of the {\it branching diagram\/} \cite{pauncz} displayed in 
Fig.\ \ref{brandiag}. The axes in this plot describe the number $N$ of fermions  
(horizontal axis) and the quantum number $S$ of the total spin (vertical axis).
At each point $(N,S)$, a circle is drawn containing the number $g(N,S)$ which
gives the degeneracy of spin states. It is found\cite{pauncz} that
\begin{equation}
g(N,S) = 
\left( \begin{array}{c} N \\ N/2-S \end{array} \right) -
\left( \begin{array}{c} N \\ N/2-S-1 \end{array} \right).
\label{degen}
\end{equation}

Specifically for $N=4$ particles, there is one spin eigenfunction with
$S=2$, three with $S=1$, and two with $S=0$. In general the spin part of the
EXD wave functions involves a linear superposition over all the degenerate spin 
eigenfunctions for a given $S$. 

For a small number of particles, one
can find compact expressions that encompass all possible superpositions. For
example, for $N=4$ and $S=0$, $S_z=0$ one has: \cite{note3}

\begin{eqnarray}
 {\cal X}_{00}&=&\sqrt{\frac{1}{3}}\sin\theta |
\uparrow\uparrow\downarrow\downarrow\rangle+(\frac{1}{2}\cos\theta-
       \sqrt{\frac{1}{12}}\sin\theta)\left|
\uparrow\downarrow\uparrow\downarrow\right> \nonumber \\
       & & -(\frac{1}{2}\cos\theta+\sqrt{\frac{1}{12}}\sin\theta)|
\uparrow\downarrow\downarrow\uparrow\rangle \nonumber \\
       & & -(\frac{1}{2}\cos\theta+\sqrt{\frac{1}{12}}\sin\theta)\left|
\downarrow\uparrow\uparrow\downarrow\right> \nonumber \\
       & & +(\frac{1}{2}\cos\theta-\sqrt{\frac{1}{12}}\sin\theta)\left|
\downarrow\uparrow\downarrow\uparrow\right>
        +\sqrt{\frac{1}{3}}\sin\theta\left|
\downarrow\downarrow\uparrow\uparrow\right>,\nonumber\\
\label{x00}
\end{eqnarray}
where the parameter $\theta$ satisfies $-\pi/2\leq \theta \leq \pi/2$ 
and is chosen such that $\theta=0$ corresponds to the spin function with 
intermediate two-electron spin $S_{12}=0$ and three-electron spin
$S_{123}=1/2$; whereas 
$\theta=\pm \pi/2$ corresponds to the one with intermediate spins $S_{12}=1$ 
and $S_{123}=1/2$.

For $N=4$ and $S=1$, $S_z=0$ one has:
\begin{eqnarray}
{\cal X}_{10} &=& \nonumber \\
&& \hspace{-1cm} (\sqrt{\frac{1}{6}}\sin\theta\sin\varphi-
\sqrt{\frac{1}{12}}\sin\theta\cos\varphi-\frac{1}{2}\cos\theta)
             \left|\downarrow\uparrow\uparrow\downarrow\right> \nonumber \\
&& \hspace{-1cm}      +(\sqrt{\frac{1}{6}}\sin\theta\sin\varphi-
\sqrt{\frac{1}{12}}\sin\theta\cos\varphi+\frac{1}{2}\cos\theta)
             \left|\uparrow\downarrow\uparrow\downarrow\right> \nonumber \\
&& \hspace{-1cm}        +(\sqrt{\frac{1}{12}}\sin\theta\cos\varphi-
\sqrt{\frac{1}{6}}\sin\theta\sin\varphi-\frac{1}{2}\cos\theta)
             \left|\downarrow\uparrow\downarrow\uparrow\right> \nonumber \\
&& \hspace{-1cm}        +(\sqrt{\frac{1}{12}}\sin\theta\cos\varphi-
\sqrt{\frac{1}{6}}\sin\theta\sin\varphi+\frac{1}{2}\cos\theta)
             \left|\uparrow\downarrow\downarrow\uparrow\right> \nonumber \\
&& \hspace{-1cm}      +(\sqrt{\frac{1}{6}}\sin\theta\sin\varphi+
\sqrt{\frac{1}{3}}\sin\theta\cos\varphi)
             \left|\uparrow\uparrow\downarrow\downarrow\right> \nonumber \\
&& \hspace{-1cm}      -(\sqrt{\frac{1}{6}}\sin\theta\sin\varphi+
\sqrt{\frac{1}{3}}\sin\theta\cos\varphi)
             \left|\downarrow\downarrow\uparrow\uparrow\right>,
\label{x10}
\end{eqnarray}
where the parameters $\theta$ and $\varphi$ satisfy $-\pi/2\leq \theta 
\leq \pi/2$ and $-\pi/2\leq \varphi \leq \pi/2$. Three independent spin 
functions with definite intermediate two-electron, $S_{12}$, and three-electron, 
$S_{123}$, spin values correspond to the $\theta$ and $\varphi$ values as 
follows: for $S_{12}=0$ and $S_{123}=1/2$, $\theta=0$; 
for $S_{12}=1$ and $S_{123}=1/2$, $\theta=\pm \pi/2$ and $\varphi=0$; and for 
$S_{12}=1$ and $S_{123}=3/2$, $\theta=\pm \pi/2$ and $\varphi=\pm \pi/2$.

Finally, for $N=4$ and $S=2$, $S_z=0$ (maximum polarization) case, one has:
\begin{eqnarray}
{\cal X}_{20} &=& \nonumber \\
&& \hspace{-1.1cm}
\frac{ \left|\downarrow\downarrow\uparrow\uparrow\right>+
\left|\downarrow\uparrow\downarrow\uparrow\right>+
       \left|\downarrow\uparrow\uparrow\downarrow\right> 
 +\left|\uparrow\downarrow\downarrow\uparrow\right>+
       \left|\uparrow\downarrow\uparrow\downarrow\right>+
\left|\uparrow\uparrow\downarrow\downarrow\right>}{\sqrt{6}}. \nonumber \\
\label{x20}
\end{eqnarray}

%
\begin{figure}[t]
\centering{
\includegraphics[width=8.4cm]{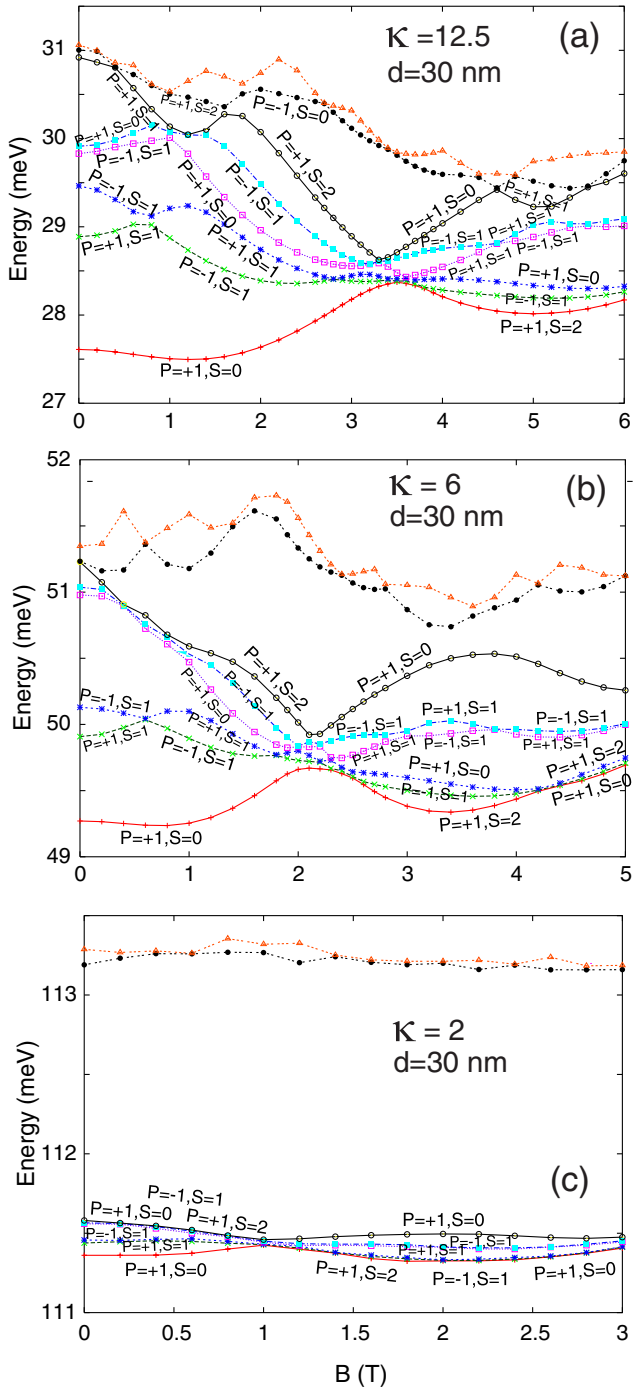}}
\caption{(Color online) Energy spectra (as a function of the magnetic field $B$) 
for $N=4$ electrons in a double quantum dot with interdot separation  $d=30$ nm.
(a) $\kappa =12.5$ corresponds to GaAs. (b) $\kappa =6$. (c) $\kappa =2$. 
The Zeeman term was neglected, and thus all states with the same total spin $S$ 
different spin projections $S_z$ are degenerate. Remaining 
parameters: $\epsilon^b=0.5$, $\hbar \omega_0=5.1$ meV, $m^*=0.07 m_e$,
$h_1=h_2=0$. For all figures in this paper, $\hbar \omega_0$, $m^*$, and
$h_1=h_2$ are the same.   
Energies are referenced to $N \hbar \sqrt{\omega_0^2 + \omega_c^2/4}$, 
where $\omega_c=eB/(m^* c)$ is the cyclotron frequency.
}
\label{spd30e05}
\end{figure}

\section{Exact-diagonalization results}
\label{exdres}

\subsection{Energy spectra}
\label{exdspec}

The excitation spectra as a function of the applied magnetic field for four 
electrons in a double QD with interdot distance $d=2x_2=-2x_1=30$ nm and 
no voltage bias between the dots [$h_1=h_2=0$, see Eq.\ (\ref{hsp})] are plotted
for three different values of the interelectron repulsion strength, i.e., weak 
[$\kappa=12.5$ (GaAs); see Fig.\ \ref{spd30e05}(a)], intermediate [$\kappa=6$; 
see Fig.\ \ref{spd30e05}(b)], and strong [$\kappa=2$; see Fig.\ 
\ref{spd30e05}(c)] Coulomb repulsion. The interdot barrier parameter was taken 
as $\epsilon^b=0.5$ (because $h_1=h_2=0$, one has 
$\epsilon^b_1=\epsilon^b_2=\epsilon^b$; see Section \ref{2dhopot} for the 
definitions). In all cases, we calculated the eight lowest energy levels.

We observe that the lowest six levels form a band that separates from the rest 
of the spectrum through the opening of a gap. This happens already at a 
relatively weak interelectron repulsion, and it is well developed for the 
intermediate case ($\kappa=6$). It is of interest to note that the number of 
levels in the band (six) coincides with
the total number of spin eigenfunctions for $N=4$ fermions, as can be seen
from the branching diagram displaying the spin degeneracies.
In particular, there is one level with total spin $S=2$ (and parity 
${\cal P}_{xy}=1$), three levels with total spin $S=1$ (two with 
${\cal P}_{xy}=1$ and one with ${\cal P}_{xy}=-1$), and two levels with total 
spin $S=0$ (one with ${\cal P}_{xy}=1$ and the second with ${\cal P}_{xy}=-1$).
All these six levels approximately ``cross'' at one point \cite{note9} situated
at about $B \approx 3.5$ T for $\kappa=12.5$, $B \approx 2.2$ T for $\kappa=6$,
and $B \approx 1.1$ T for $\kappa=2$.

%
\begin{figure}[t]
\centering{
\includegraphics[width=8.4cm]{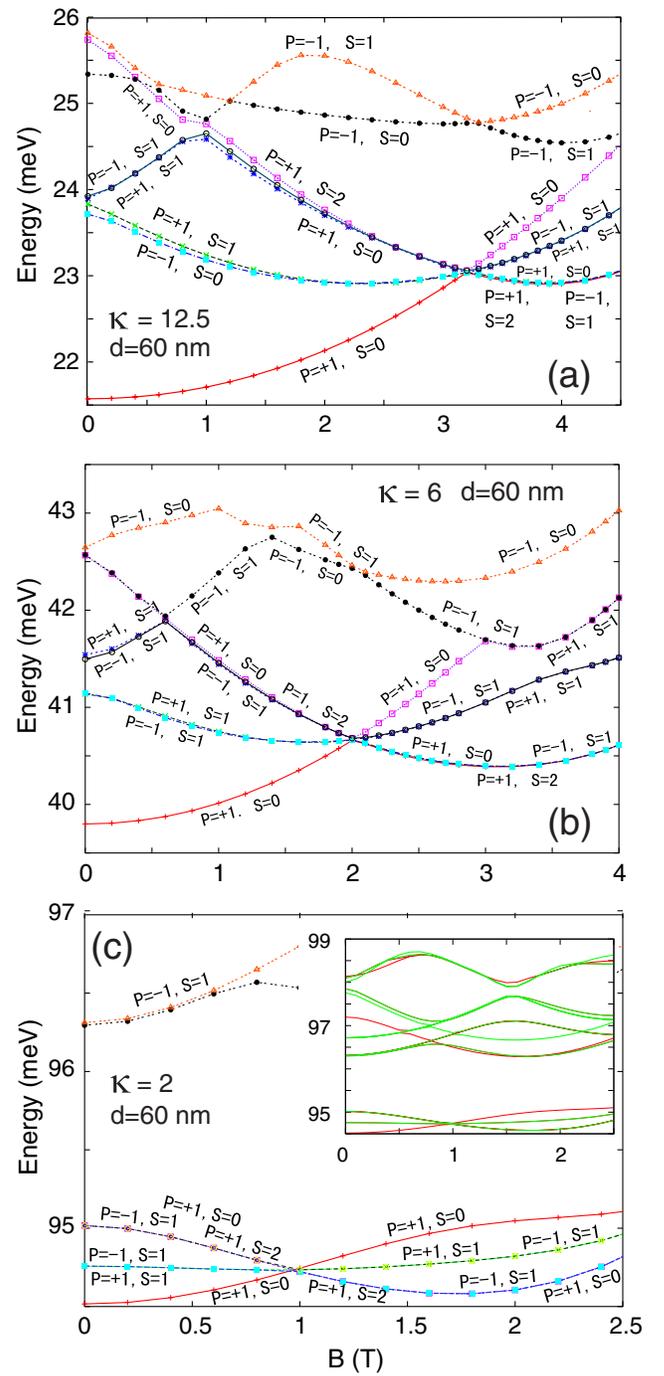}}
\caption{(Color online) Energy spectra (as a function of the magnetic field $B$) for
$N=4$ electrons in a double quantum dot with interdot separation  $d=60$ nm.
(a) $\kappa =12.5$ (weak interelectron repulsion) corresponds to GaAs. 
(b) $\kappa =6$. (c) $\kappa =2$. 
The interdot barrier corresponds to $\epsilon^b=0.5$.
The inset in (c) displays all the lowest 12 $P_{xy}=1$ and the lowest 10 
$P_{xy}=-1$ energies, and it demonstrates a trend toward formation of higher 
bands. Energies are referenced to $N \hbar \sqrt{\omega_0^2 + \omega_c^2/4}$,
where $\omega_c=eB/(m^* c)$ is the cyclotron frequency.
}
\label{spd60e05}
\end{figure}

The trends associated with the opening of a gap and the formation of a
six-state low band appear further reinforced 
for the larger interdot distance of $d=60$ nm (displayed in Fig.\ 
\ref{spd60e05} for the three values of the dielectric
constant $\kappa=12.5$, 6, and 2, respectively). It is remarkable that the 
six lower curves ``cross'' now at a sharply defined point \cite{note9} 
(situated at $B \approx 3.3$ T for $\kappa=12.5$, $B \approx 2.1$ T for 
$\kappa=6$, and $B \approx 1.0$ T for $\kappa=2$). 
The six curves demonstrate additional near degeneracies regrouping
approximately to three curves before and after the crossing point, which
results in a remarkable simplification of the spectrum.

For strong repulsion ($\kappa=2$), all six states in the low band are 
almost degenerate for both distances [$d=30$ nm; see Fig.\ 
\ref{spd30e05}(c) and $d=60$ nm; see Fig.\ \ref{spd60e05}(c)]. This is a 
consequence of the formation of a near-rigid Wigner molecule (WM) with strongly 
localized electrons. Namely, the overlaps between the orbitals of localized
electrons are very small (see, e.g., Ref.\ \onlinecite{yann99}), 
yielding small exchange contributions in the total energies,\cite{yann02} and 
thus all six possible spin multiplicities tend to become degenerate in energy. 
Furthermore the physical picture of a near-rigid Wigner molecule suggests that 
the energy gap to the next group of states corresponds to excitation of the 
lowest stretching vibrational mode of the 4-electron molecule.

Since the main panels in Figs.\ \ref{spd30e05} and \ref{spd60e05} display only 
the four lowest-in-energy states with positive parity and the four corresponding
states with negative parity, one needs a larger part of the spectrum to 
ascertain whether higher bands are formed. To this end, the inset in Fig.\ 
\ref{spd60e05}(c) displays the twelve lowest-in-energy curves with $P_{xy}=1$ 
and the ten corresponding curves with $P_{xy}=-1$. In addition to the 
6-state low band, the inset indicates formation of a higher band 
comprising a total of 12 states (not labeled); however, a detailed study of this
higher band falls outside the scope of the present paper.

It is natural to anticipate at this point that the above behavior of the
low-energy EXD spectra at low $B$ can be generalized to an arbitrary number of 
electrons $N$ in a double QD. Namely,
as the strength of the interelectron interaction increases, a low-energy band
comprising all possible spin multiplicities will form and it will become
progressively well separated by an energy gap from the higher excitations.
For example, for $N=6$, an inspection of the branching diagram in Fig.\
\ref{brandiag} leads us to the prediction that there will be 20 states in this
low-energy band. A similar behavior emerges also in the case of a {\it single\/},
but strongly {\it anisotropic\/} quantum dot; indeed a low-energy band of three 
states (see the branching diagram in Fig.\ \ref{brandiag}) has been found for 
$N=3$ electrons in Ref.\ \onlinecite{yues07}.

It is of interest to contrast the above behavior of the excitation spectra in a 
double QD with that of an $N$-electron circular dot. Specifically, in the 
circular QD, large inetelectron repulsion leads to formation of a near-rigid 
{\it rotating\/} Wigner molecule that exhibits a rigid moment of inertia.
Then the states inside the low-energy band (two states for $N=2$, three for
$N=3$, six for $N=4$, etc.) do not become degenerate in energy, but form
an yrast rotational band \cite{yann04} specified by $L^2/2{\cal J}_0$, where 
$L$ is the total angular momentum and ${\cal J}_0$ is the classical moment of 
inertia. We note that the energy splittings among the yrast rotational 
states are much smaller than the vibrational energy gap in circular dots 
associated with the quantum of energy $\sqrt{3} \hbar \omega_0$ of the 
stretching (often referred to as breathing) mode
of the polygonal-ring configuration of the quasiclassical Wigner molecule.
\cite{yann00,peet95,vign96}

\subsection{Electron densities}
\label{eldensec}

The electron density is the expectation value of the one-body operator 
\begin{equation} 
\hat{\rho}({\bf r})= \sum_{i=1}^N \delta({\bf r}-{\bf r}_i),
\label{eldenop}
\end{equation}
that is:
\begin{eqnarray}
\rho({\bf r}) &=& \langle \Phi^{\text{EXD}}_{N,q} 
\vert \hat{\rho}({\bf r}) \vert \Phi^{\text{EXD}}_{N,q} \rangle \nonumber \\
&=& \sum_{I,J} C_I^{q*} C_J^q 
\langle \Psi_I^N \vert \hat{\rho}({\bf r}) \vert \Psi_J^N \rangle.
\label{elden}
\end{eqnarray}

Since $\hat{\rho}({\bf r})$ is a one-body operator, it connects only Slater 
determinants $\Psi_I^N$ and $\Psi_J^N$ that differ at most by {\it one\/} spin 
orbital $\chi_j({\bf r})$; for the corresponding Slater rules for calculating 
matrix elements between determinants for one-body operators in terms of spin 
orbitals, see Table 2.3 in Ref.\ \onlinecite{szabbook}.

\begin{figure}[t]
\centering{
\includegraphics[width=8cm]{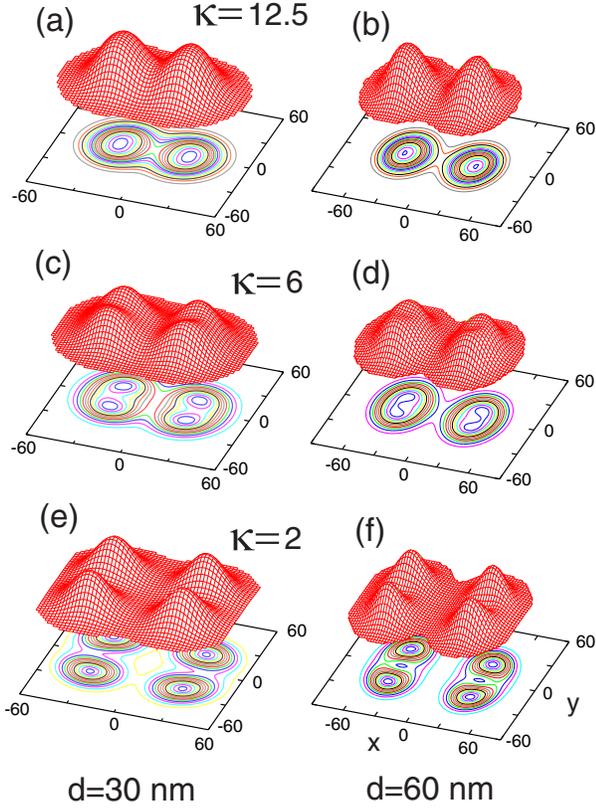}}
\caption{(Color online) Electron densities at $B=0$ for the ground state (with 
$S=0$, $S_z=0$, parity $P_{xy}=1$) for $N=4$ electrons in a double quantum dot with 
interdot separations  $d=30$ nm (left column) and $d=60$ nm (right column). 
The top (a-b), middle (c-d), and bottom (e-f) rows correspond to 
$\kappa =12.5$ (weak repulsion), $\kappa=6$ (intermediate repulsion), and
$\kappa=2$ (strong repulsion), respectively. 
Ground-state energies: (a) $E=27.609$ meV, (b) $E=21.572$ meV,
(c) $E=49.217$ meV, (d) $E=39.799$ meV, (e) $E=111.361$ meV, (f) $E=94.516$ meV
(compare Figs. \ref{spd30e05}$-$\ref{spd60e05}).
The interdot barrier corresponds to $\epsilon^b=0.5$. Distances in nm.
Vertical axis in arbitrary units (with the same scale for all six panels).
}
\label{elden30_60}
\end{figure}

In Figs.\ \ref{elden30_60}(a-f), we display (for the aforementioned three 
strengths of interelectron repulsion) the ground-state electron
densities for for $N=4$ electrons in the case of a double dot at zero
magnetic field with interdot separations $d=30$ nm (left column) and
$d=60$ nm (right column). 

%

For the weak interaction case ($\kappa=12.5$) at $B=0$, the electron densities 
do not exhibit clear signatures of formation of a Wigner molecule for either
interdot distance, $d=30$ nm [Fig.\ \ref{elden30_60}(a)] or $d=60$ nm
[Fig.\ \ref{elden30_60}(b)]. The Wigner molecule is well formed, however,
in the case of the intermediate Coulomb repulsion [$\kappa=6$; see Figs.\ 
\ref{elden30_60}(c-d)]. One observes 
indeed four humps that correspond to the four localized electrons; they are 
located at ($\pm$34.88 nm, $\pm$13.13 nm) in the $d=60$ nm case. 
In the case of strong Coulomb repulsion ($\kappa=2$) and for the same interdot 
distance $d=60$ nm, the electrons are 
further localized as can be seen from Fig.\ \ref{elden30_60}(f);
the four humps occur now at ($\pm$39.86 nm, $\pm$21.02 nm). 
The Wigner molecule is also well formed in the
the strong-repulsion and $d=30$ nm case, as can be seen from Fig.\
\ref{elden30_60}(e), with the localized electrons located at 
($\pm$29.28 nm, $\pm$21.11 nm).

\subsection{Spin-resolved conditional probability distributions}
\label{cpdsb0}

\subsubsection{Definitions}

In the regime corresponding to a well-defined Wigner molecule, the electron
densities (see Sect. \ref{eldensec}) are characterized by four humps that
reflect the localization of the four electrons in the double quantum dot.
Such charge densities do not provide any information concerning the spin
structure of each EXD state. In fact, all six EXD states in the lower band
exhibit very similar four-humped electron densities.

The spin configurations associated with a given $(S,S_z)$ EXD state in 
the WM regime can be explored with the help of the spin-resolved
two-point anisotropic correlation function defined as:
%
\begin{figure}[t]
\centering{
\includegraphics[width=8cm]{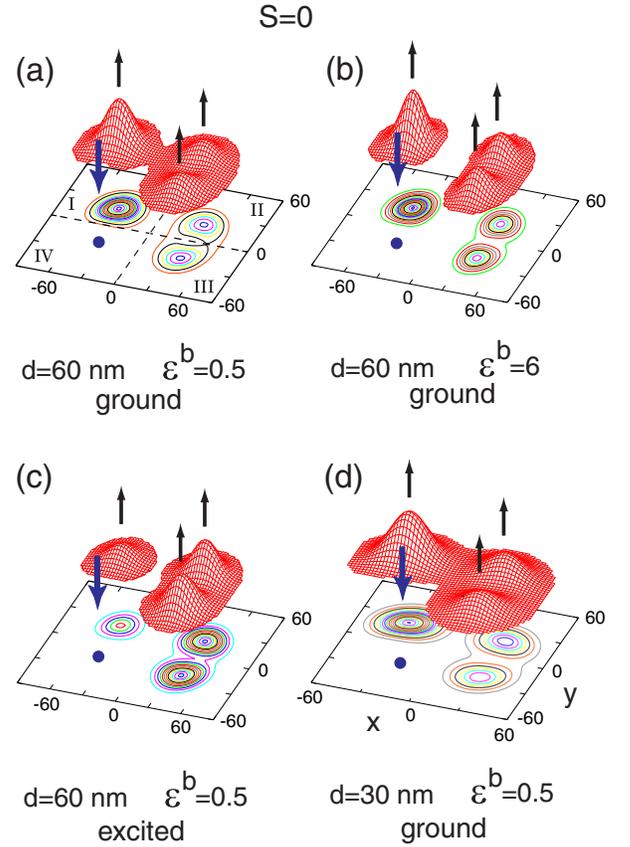}}
\caption{(Color online) CPDs ${\cal P}_{\uparrow\downarrow}$ at $B=0$ for several 
EXD states with $S=0, S_z=0$, and parity $P_{xy}=1$ of $N=4$ electrons in a double 
quantum dot with interdot separations $d=60$ nm (a-c) and $d=30$ nm (d). 
Case of strong Coulomb repulsion ($\kappa =2$) with an interdot barrier 
$\epsilon^b=0.5$ (a,c-d) and $\epsilon^b=6$ (b).
Panels (a-b,d) correspond to ground states. Panel (c) corresponds to 
the excited second $S=0$ state for the same parameters as in panel (a)
[see Fig.\ \ref{spd60e05}(c) and the branching diagram in Fig.\ \ref{brandiag}].
Energies: (a) $E=94.516$ meV, (b) $E=96.811$ meV, (c) $E=95.017$ meV, and
(d) $E=111.361$ meV [compare Figs. \ref{spd30e05}(c) and 
\ref{spd60e05}(c)].
Distances in nm. Vertical axis in arbitrary units 
(with the same scale for all panels in Figs.\ \ref{s0b0} $-$
\ref{s0b2t}). The fixed point is located at the maximum of the hump in 
the lower-left quadrant of the corresponding electron density, i.e., at
${\bf r}_0=$($-$40 nm, $-$21 nm) for panels (a-c) and 
${\bf r}_0=$($-$29 nm, $-$19 nm) for panel (d).
}
\label{s0b0}
\end{figure}

\begin{eqnarray}
&& \hspace{-0.8cm} P_{\sigma\sigma_0}({\bf r}, {\bf r}_0)=  \nonumber \\
&& \hspace{-0.5cm} \langle \Phi^{\text{EXD}}_{N,q} |
\sum_{i \neq j} \delta({\bf r} - {\bf r}_i) \delta({\bf r}_0 - {\bf r}_j)
\delta_{\sigma \sigma_i} \delta_{\sigma_0 \sigma_j}
|\Phi^{\text{EXD}}_{N,q}\rangle,
\label{tpcorr}
\end{eqnarray}
with the EXD many-body wave function given by equation (\ref{mbwf}).

Using a normalization constant
\begin{equation}
{\cal N}(\sigma,\sigma_0,{\bf r}_0) = 
\int P_{\sigma\sigma_0}({\bf r}, {\bf r}_0) d{\bf r},
\label{norm}
\end{equation}
we further define a related conditional probability distribution (CPD) as
\begin{equation}
{\cal P}_{\sigma\sigma_0}({\bf r}, {\bf r}_0) =
P_{\sigma\sigma_0}({\bf r}, {\bf r}_0)/{\cal N}(\sigma,\sigma_0,{\bf r}_0),
\label{cpd}
\end{equation}
having the property 
$\int {\cal P}_{\sigma\sigma_0}({\bf r}, {\bf r}_0) d{\bf r} =1$.
The spin-resolved CPD gives the spatial probability distribution of 
finding a second electron with spin projection $\sigma$ under the 
condition that another electron is located (fixed) at ${\bf r}_0$ with spin 
projection $\sigma_0$; $\sigma$ and $\sigma_0$ can be either up $(\uparrow$) or
down ($\downarrow$). 

To calculate $P_{\sigma\sigma_0}({\bf r}, {\bf r}_0)$ in Eq.\ (\ref{tpcorr}),
we use a symmetrized operator
\begin{eqnarray}
&& \hspace{-0.3cm} \hat{T}_{\sigma\sigma_0}({\bf r}, {\bf r}_0) = \nonumber \\
&& \sum_{i<j} \left[
\delta({\bf r} - {\bf r}_i) \delta({\bf r}_0 - {\bf r}_j)
\delta_{\sigma \sigma_i} \delta_{\sigma_0 \sigma_j} + \right. \nonumber \\
&& \hspace{0.8cm} \left. 
\delta({\bf r} - {\bf r}_j) \delta({\bf r}_0 - {\bf r}_i)
\delta_{\sigma \sigma_j} \delta_{\sigma_0 \sigma_i} \right],
\label{sympop}
\end{eqnarray}
yielding
\begin{eqnarray}
P_{\sigma\sigma_0}({\bf r}, {\bf r}_0) &=& \langle \Phi^{\text{EXD}}_{N,q} 
\vert \hat{T} \vert \Phi^{\text{EXD}}_{N,q} \rangle \nonumber \\
&=& \sum_{I,J} C_I^{q*} C_J^q 
\langle \Psi_I^N \vert \hat{T} \vert \Psi_J^N \rangle.
\label{psymm}
\end{eqnarray}

Since  $ \hat{T}_{\sigma\sigma_0}({\bf r}, {\bf r}_0)$ 
is a two-body operator, it connects only Slater determinants $\Psi_I^N$ 
and $\Psi_J^N$ that differ at most by {\it two\/} spin orbitals
$\chi_{j_1}({\bf r})$ and $\chi_{j_2}({\bf r})$; for the corresponding Slater 
rules for calculating matrix elements between determinants for two-body 
operators in terms of spin orbitals, see Table 2.4 in Ref.\ 
\onlinecite{szabbook}.

\subsubsection{Examples of $S=0$, $S_z=0$ EXD states at $B=0$}

For each charge density corresponding to a given state of the system, 
one can plot four different spin-resolved CPDs, i.e.,
${\cal P}_{\uparrow\uparrow}$, ${\cal P}_{\uparrow\downarrow}$,
${\cal P}_{\downarrow\uparrow}$, and ${\cal P}_{\downarrow\downarrow}$.
This can potentially lead to a very large number of time consuming computations
and an excessive number of plots. For studying the spin structure of the 
$S=0, S_z=0$ states at $B=0$, however, we found that knowledge of a single CPD, 
taken here to be ${\cal P}_{\uparrow\downarrow}$ (see Fig.\ \ref{s0b0}),
is sufficient in the regime of Wigner-molecule formation.
Indeed, the specific angle $\theta$ specifying the spin function ${\cal X}_{00}$
Eq.\ (\ref{x00}) corresponding to the CPDs portrayed in Fig.\ \ref{s0b0}
can be determined through the procedure described in the following:

We designate with roman indices $I$, $II$, $III$, and $IV$ the four quadrants of
the $(x,y)$ plane, starting with the upper left quadrant and going clockwise 
[see Fig.\ \ref{s0b0}(a)]. In the case of a 4$e$ 
Wigner-molecule, a single electron is localized within each quadrant. The
same roman indices designate also the positions of the localized electrons in
each of the six Slater determinants (e.g., 
$|\uparrow\uparrow\downarrow\downarrow\rangle$,
$|\uparrow\downarrow\uparrow\downarrow\rangle$, etc.) that enter into the spin 
function ${\cal X}_{00}$ in Eq.\ (\ref{x00}). We take always the fixed point to
correspond to the fourth $(IV)$ quadrant [bottom left in Fig.\ \ref{s0b0}(a)].
An inspection of Eq.\ (\ref{x00}) shows that only three Slater determinants in 
${\cal X}_{00}$ contribute to ${\cal P}_{\uparrow\downarrow}$, namely  
$|\uparrow\uparrow\downarrow\downarrow\rangle$, 
$|\uparrow\downarrow\uparrow\downarrow\rangle$, and
$|\downarrow\uparrow\uparrow\downarrow\rangle$; these are the only
determinants in Eq.\ (\ref{x00}) with a down spin in the 4th quadrant.
From these three Slater determinants, only the first and the second
contribute to the conditional probability $\Pi_{\uparrow\downarrow}(I)$ of 
finding another electron with spin-up in quadrant $I$; this corresponds to the
volume under the hump of the EXD CPD in quadrant $I$ [see, e.g., the hump in 
Fig.\ \ref{s0b0}(a)]. Taking the squares of the coefficients of
$|\uparrow\uparrow\downarrow\downarrow\rangle$ and 
$|\uparrow\downarrow\uparrow\downarrow\rangle$ in Eq.\ (\ref{x00}), one
gets 
\begin{equation}
\Pi_{\uparrow\downarrow}(I) \propto 
\frac{\sin^2 \theta}{3} +  
\left( \frac{1}{2} \cos \theta - \sqrt{ \frac{1}{12} } \sin \theta \right)^2.
\label{pi_updown_i}
\end{equation} 

Similarly, one finds that only 
$|\uparrow\uparrow\downarrow\downarrow\rangle$ and
$|\downarrow\uparrow\uparrow\downarrow\rangle$ contribute to
$\Pi_{\uparrow\downarrow}(II)$, and that
\begin{equation}
\Pi_{\uparrow\downarrow}(II) \propto
\frac{\sin^2 \theta}{3} +
\left( \frac{1}{2} \cos \theta + \sqrt{ \frac{1}{12} } \sin \theta \right)^2.
\label{pi_updown_ii}
\end{equation}

Integrating under the humps of the EXD CPD in quadrants $I$ and $II$, we 
determine numerically the ratio 
$\Pi_{\uparrow\downarrow}(I)/\Pi_{\uparrow\downarrow}(II)$, which allows
us to specify the absolute value of $\theta$ (within the interval
$-90^\circ \leq \theta \leq 90^\circ$) via the expressions in Eqs.\ 
(\ref{pi_updown_i}) and (\ref{pi_updown_ii}).  The restriction to the
absolute value of $\theta$ is a result of the squares of the sine and cosine 
entering in $\Pi_{\uparrow\downarrow}(I)$ and $\Pi_{\uparrow\downarrow}(II)$. 
To obtain the actual sign of $\theta$, additional information is needed:
for example the ratio $\Pi_{\uparrow\downarrow}(I)/\Pi_{\uparrow\downarrow}(III)$
can be used in a similar way, where
\begin{eqnarray}
\Pi_{\uparrow\downarrow}(III) & \propto &
\left( \frac{1}{2} \cos \theta - \sqrt{ \frac{1}{12} } \sin \theta \right)^2 +
\nonumber \\
&& \left( \frac{1}{2} \cos \theta + \sqrt{ \frac{1}{12} } \sin \theta \right)^2.
\label{pi_updown_iii}
\end{eqnarray}

Using the method described above, we find that $\theta \approx -60^\circ$ for
the EXD ground state at $d=60$ nm (larger interdot distance) and $\kappa=2$ 
[strong repulsion; see Fig.\ \ref{s0b0}(a)], and the corresponding spin 
function simplifies to
\begin{equation}
{\cal X}_{00}^{(1)}=-\frac{1}{2} |\uparrow\uparrow\downarrow\downarrow\rangle
+\frac{1}{2} |\uparrow\downarrow\uparrow\downarrow\rangle
+\frac{1}{2} |\downarrow\uparrow\downarrow\uparrow\rangle
-\frac{1}{2} |\downarrow\downarrow\uparrow\uparrow\rangle.
\label{x00_cpd60k2b0l1}
\end{equation}

Remarkably, increasing the interdot barrier from $\epsilon^b=0.5$ 
[Fig.\ \ref{s0b0}(a)] to $\epsilon^b=6$ [Fig.\ \ref{s0b0}(b)], while 
keeping the other parameters constant, does not influence much the composition
of the associated spin function, which remains that given by Eq.\
(\ref{x00_cpd60k2b0l1}). This happens in spite of the visible change in the
degree of localization in the electronic orbitals, with the higher 
interdot-barrier case exhibiting a sharper localization.  

In Fig.\ \ref{s0b0}(c), we display the ${\cal P}_{\uparrow\downarrow}$ CPD
for an excited state with $S=0, S_z=0$ (having $P_{xy}=1$ and energy
$E=95.017$ meV), with the remaining parameters being the same as in Fig.\
\ref{s0b0}(a). For this case, following an analysis as described above,
we found the angle $\theta \approx 30^\circ$, which is associated with
a spin function of the form
\begin{eqnarray}
{\cal X}_{00}^{(2)} &=& 
\frac{1}{2\sqrt{3}} |\uparrow\uparrow\downarrow\downarrow\rangle
+\frac{1}{2\sqrt{3}} |\uparrow\downarrow\uparrow\downarrow\rangle
-\frac{1}{\sqrt{3}} |\uparrow\downarrow\downarrow\uparrow\rangle \nonumber \\
&& -\frac{1}{\sqrt{3}} |\downarrow\uparrow\uparrow\downarrow\rangle
+\frac{1}{2\sqrt{3}} |\downarrow\uparrow\downarrow\uparrow\rangle
+\frac{1}{2\sqrt{3}} |\downarrow\downarrow\uparrow\uparrow\rangle.\nonumber \\
\label{x00_cpd60k2b0l3}
\end{eqnarray}

We note that the  spin functions in Eqs.\ (\ref{x00_cpd60k2b0l1}) and 
(\ref{x00_cpd60k2b0l3}) are orthogonal.

In Fig.\ \ref{s0b0}(d), we display the ${\cal P}_{\uparrow\downarrow}$ CPD
for the ground state with $S=0, S_z=0$ (having $P_{xy}=1$ and energy
$E=111.361$ meV) and for the shorter interdot distance $d=30$ nm.
For this case, we found an angle $\theta \approx -63.08^\circ$, which 
corresponds to the following spin function:
\begin{eqnarray}
{\cal X}_{00}^{(3)} &=& 
-0.5148 |\uparrow\uparrow\downarrow\downarrow\rangle
+0.4838 |\uparrow\downarrow\uparrow\downarrow\rangle
+0.031 |\uparrow\downarrow\downarrow\uparrow\rangle \nonumber \\
&& 0.031 |\downarrow\uparrow\uparrow\downarrow\rangle
+0.4838 |\downarrow\uparrow\downarrow\uparrow\rangle
-0.5148 |\downarrow\downarrow\uparrow\uparrow\rangle. \nonumber \\
\label{x00_cpd30k2b0l1}
\end{eqnarray}

From a comparison of the above result with that for the larger 
$d=60$ nm [see Eq.\ (\ref{x00_cpd60k2b0l1})], we conclude 
that the difference in interdot distance results in a slight variation
of the spin functions.

\subsubsection{Examples of $S=1$, $S_z=0$ EXD states at $B=0$}
\label{sec_s_one}

\begin{figure}[t]
\centering{
\includegraphics[width=8.0cm]{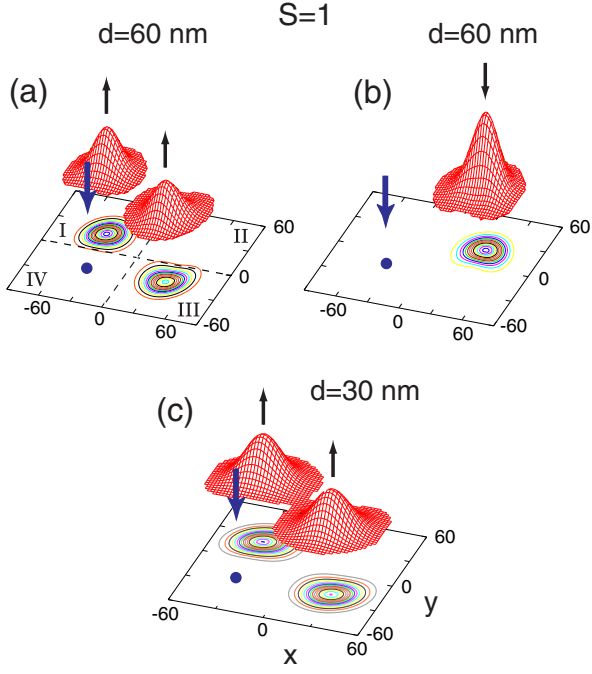}} 
\caption{(Color online) CPDs at $B=0$ for excited EXD states with $S=1, S_z=0$, and 
parity $P_{xy}=1$ of $N=4$  electrons in a double quantum dot at the larger interdot 
separation $d=60$ nm (a-b) and the shorter interdot separation $d=30$ nm (c).
Panels (a) and (c) display ${\cal P}_{\uparrow\downarrow}$ CPDS (down-up), 
while panel (b) displays a different ${\cal P}_{\downarrow\downarrow}$ CPD
(down-down), but for the {\it same\/} state as in (a). 
Case of strong Coulomb repulsion 
($\kappa =2$) with interdot barrier $\epsilon^b=0.5$. Energies: (a-b) 
$E=94.757$ meV, and (c) $E=111.438$ meV [compare Figs. \ref{spd30e05}(c) and
\ref{spd60e05}(c)]. Distances in nm. Vertical axis in arbitrary units
(with the same scale for all panels in Figs.\ \ref{s0b0} $-$
\ref{s0b2t}). The fixed point is located at the maximum of the hump in 
the lower-left quadrant of the corresponding electron density, i.e., at
${\bf r}_0=$($-$40 nm, $-$21 nm) for panels (a-b) and  
${\bf r}_0=$($-$29 nm, $-$19 nm) for panel (c). Note that this is a case with 
$S=1$; the previous Fig.\ \ref{s0b0} displayed $S=0$ cases.
}
\label{s1b0}
\end{figure}

In this section, we turn our attention to partially polarized EXD states with
$S=1$. 

In Fig.\ \ref{s1b0}(a), we display the ${\cal P}_{\uparrow\downarrow}$ CPD 
at $B=0$ for an excited state with $S=1, S_z=0$, parity $P_{xy}=1$, and energy 
$E=94.757$ meV, at the larger 
interdot separation $d=60$ nm. Again we consider the case of strong Coulomb 
repulsion ($\kappa =2$) with an interdot barrier $\epsilon^b=0.5$.
The corresponding spin function ${\cal X}_{10}$ [Eq.\ (\ref{x10})] 
depends on two different angles $\theta$ and $\phi$, and one needs at least
two different CPDs for determining their specific values. For this purpose,
we display also the ${\cal P}_{\downarrow\downarrow}$ CPD for the same state in
Fig.\ \ref{s1b0}(b).

The specific values of $\theta$ and $\phi$ associated with the CPDs in
Figs.\ \ref{s1b0}(a) and \ref{s1b0}(b) can be determined through
the ratios $\Pi_{\uparrow\downarrow}(I)/\Pi_{\uparrow\downarrow}(II)$
and $\Pi_{\uparrow\downarrow}(I)/\Pi_{\uparrow\downarrow}(III)$
[associated with Fig.\ \ref{s1b0}(a)] and 
$\Pi_{\downarrow\downarrow}(I)/\Pi_{\downarrow\downarrow}(II)$
and $\Pi_{\downarrow\downarrow}(I)/\Pi_{\downarrow\downarrow}(III)$
[associated with Fig.\ \ref{s1b0}(b)], where
\begin{eqnarray}
\Pi_{\uparrow\downarrow}(I)& \propto&
 \frac{1}{3} \sin^2 \theta \sin^2 \phi
+\frac{5}{12} \sin^2 \theta \cos^2 \phi  
+ \frac{1}{4} \cos^2 \theta \nonumber \\
&& \hspace{-0.5cm} + \frac{\sqrt{2}}{6} \sin^2 \theta \sin \phi \cos \phi 
+ \frac{1}{\sqrt{6}} \sin \theta \cos \theta \sin \phi \nonumber \\
&& \hspace{-0.5cm} -\frac{1}{\sqrt{12}} \sin \theta \cos \theta \cos \phi, 
\label{pi_x10_1_i}
\end{eqnarray} 

\begin{eqnarray}
\Pi_{\uparrow\downarrow}(II) &\propto&
 \frac{1}{3} \sin^2 \theta \sin^2 \phi
+\frac{5}{12} \sin^2 \theta \cos^2 \phi  
+ \frac{1}{4} \cos^2 \theta \nonumber \\
&& \hspace{-0.5cm} + \frac{\sqrt{2}}{6} \sin^2 \theta \sin \phi \cos \phi 
- \frac{1}{\sqrt{6}} \sin \theta \cos \theta \sin \phi \nonumber \\
&& \hspace{-0.5cm} +\frac{1}{\sqrt{12}} \sin \theta \cos \theta \cos \phi, 
\label{pi_x10_1_ii}
\end{eqnarray} 

\begin{eqnarray}
\Pi_{\uparrow\downarrow}(III) &\propto&
 \frac{1}{3} \sin^2 \theta \sin^2 \phi
+\frac{1}{6} \sin^2 \theta \cos^2 \phi \nonumber \\  
&& \hspace{-0.5cm} -\frac{\sqrt{2}}{3} \sin^2 \theta \sin \phi \cos \phi 
+\frac{1}{2} \cos^2 \theta, 
\label{pi_x10_1_iii}
\end{eqnarray} 
and
\begin{equation}
\Pi_{\downarrow\downarrow}(I) \propto
\left( \sqrt{\frac{1}{6}} \sin \theta \sin \phi
- \sqrt{\frac{1}{12}} \sin \theta \cos \phi  
- \frac{1}{2} \cos \theta \right)^2,
\label{pi_x10_2_i}
\end{equation} 

\begin{equation}
\Pi_{\downarrow\downarrow}(II) \propto
\left( \sqrt{\frac{1}{6}} \sin \theta \sin \phi
- \sqrt{\frac{1}{12}} \sin \theta \cos \phi  
+ \frac{1}{2} \cos \theta \right)^2,
\label{pi_x10_2_ii}
\end{equation} 

\begin{equation}
\Pi_{\downarrow\downarrow}(III) \propto
\left( \sqrt{\frac{1}{6}} \sin \theta \sin \phi
+ \sqrt{\frac{1}{3}} \sin \theta \cos \phi \right)^2.
\label{pi_x10_2_iii}
\end{equation} 

Using Eqs.\ (\ref{pi_x10_1_i}) $-$ (\ref{pi_x10_2_iii}) and the numerical values
of the ratios 
$\Pi_{\uparrow\downarrow}(I)/\Pi_{\uparrow\downarrow}(II)$
and $\Pi_{\uparrow\downarrow}(I)/\Pi_{\uparrow\downarrow}(III)$ 
$\Pi_{\downarrow\downarrow}(I)/\Pi_{\downarrow\downarrow}(II)$
and $\Pi_{\downarrow\downarrow}(I)/\Pi_{\downarrow\downarrow}(III)$ (specified
via a volume integration under the humps of the EXD CPDs), we determined
that $\theta=-45^\circ$ and $\sin \phi=-\sqrt{2/3}$, $\cos \phi =\sqrt{1/3}$
(i.e., $\phi \approx -54.736^\circ$). Thus, the corresponding spin function 
reduces to the simple form
\begin{equation}
{\cal X}_{10}= 
\sqrt{\frac{1}{2}} |\uparrow\downarrow\uparrow\downarrow\rangle 
- \sqrt{\frac{1}{2}} |\downarrow\uparrow\downarrow\uparrow\rangle.
\label{x10_cpd60k2b0l2}
\end{equation} 

In Fig.\ \ref{s1b0}(c), we display the ${\cal P}_{\uparrow\downarrow}$ CPD 
at $B=0$ for a similar excited state as in Fig.\ \ref{s1b0}(a) (with $S=1, 
S_z=0$, parity $P_{xy}=1$, and energy $E=111.438$ meV) of $N=4$  electrons at 
the shorter interdot separation $d=30$ nm. Here too we consider the
case of strong Coulomb repulsion ($\kappa =2$) with interdot barrier
$\epsilon^b=0.5$. We note that the localization of electrons is stronger for
the larger interdot distance [compare Fig.\ \ref{s1b0}(a) with
Fig.\ \ref{s1b0}(c)]. This difference, however, does not influence the
coefficients entering into the associated spin function, which we found
to remain very close to the specific form in Eq.\ (\ref{x10_cpd60k2b0l2}).

\subsubsection{Spin-resolved conditional probability distributions at 
$B \neq0$}
\label{cpdsbt}

\begin{figure}[t]
\centering{
\includegraphics[width=8.0cm]{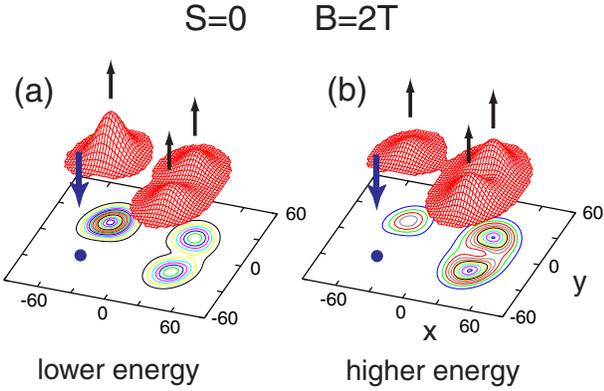}} 
\caption{(Color online) ${\cal P}_{\uparrow\downarrow}$ CPDs for the EXD states 
at $B=2$ T with $S=0$, $S_z=0$, and parity $P_{xy}=1$ 
of $N=4$ electrons in a double quantum dot at the larger interdot separation 
$d=60$ nm. (a) The lower energy of the two $S=0$ states (see branching
diagram in Fig.\ \ref{brandiag}). (b) Higher energy $S=0$ state.
Case of strong Coulomb repulsion ($\kappa =2$) with interdot barrier
$\epsilon^b=0.5$. Energies: (a) $E=94.605$ meV and (b) $E=95.047$ meV
[compare Fig. \ref{spd60e05}(c)]. Distances in nm. 
Vertical axis in arbitrary units (with the same scale for all panels in 
Figs.\ \ref{s0b0} $-$ \ref{s0b2t}). The fixed point is located at 
${\bf r}_0=$($-$40 nm, $-$21 nm). 
}
\label{s0b2t}
\end{figure}

In Fig.\ \ref{s0b2t} we display EXD CPDs at 
a finite value of the magnetic field, and precisely at $B=2$ T, for the
two states of the low-energy band with $S=0$, $S_z=0$ (at the larger interdot 
separation $d=60$ nm and strong interelectron repulsion $\kappa=2$). 
This value of $B$ was chosen to lie beyond the crossing point for the six states
of the low-energy band [which happens at $B \sim 1$ T; see 
Fig.\ \ref{spd60e05}(c)]. Comparison with the CPDs of the corresponding states
at zero magnetic field [see Figs.\ \ref{s0b0}(a) and \ref{s0b0}(c)]
shows that the spin structure of the associated Wigner molecule varies rather
slowly with the increasing magnetic field in the range $0 \leq B \leq 2.5$ T. 

Following the height of the humps in the left upper quadrants, one
observes that the CPD in Fig.\ \ref{s0b2t}(a) (case of {\it lower-energy\/}
state at $B=2$ T with $S=0$ and $P_{xy}=1$) corresponds to that of 
Fig.\ \ref{s0b0}(a) (case of {\it lower-energy\/} state at $B=0$ 
with $S=0$ and $P_{xy}=1$). Similarly, the CPD in Fig.\ \ref{s0b2t}(b) at 
$B=2$ T ({\it higher-energy\/} state) corresponds to that of Fig.\ 
\ref{s0b0}(c) at $B=0$ ({\it higher-energy\/} state). From these results, we 
concludes that the two states with $S=0$ and $P_{xy}=1$ do not really cross at 
the 'crossing' point at $B \sim 1$ T. In reality, this point is an anticrossing 
point for these two states, although the anticrossing gap is too small to be 
seen with the naked eye. This behavior agrees with that expected from states 
having the same quantum numbers. We checked that a similar observation applies
for the two other states in the low-energy band having the same quantum numbers,
i.e., those having $S=1$ and $P_{xy}=-1$.

\section{Analogies with a 4-site Heisenberg spin cluster}
\label{heis}

In Section \ref{cpdsb0}, using the spin-resolved CPDs, 
we showed that the EXD many-body 
wave functions in the Wigner-molecule regime can be expressed as a linear 
superposition of a small number of Slater determinants and that this 
superposition exhibits the structure expected from the theory of many-body spin 
functions. This finding naturally suggests a strong analogy with the field of 
nanomagnets and quantum magnetism, usually studied via the 
explicitly spin-dependent model effective Hamiltonian known as the Heisenberg 
Hamiltonian, \cite{hend93,haas07,mano91} given by:
${\cal H}^\prime_H = \sum_{i,j} J_{ij} {\bf S}_i {\bf \cdot S}_j$,
where $J_{ij}$ are the exchange integrals between spins on sites $i$ and $j$.
Even in its more familiar, simpler form 
\begin{equation}
{\cal H}_H = \sum_{\langle i,j \rangle} J_{ij} {\bf S}_i {\bf \cdot S}_j,
\label{hh2} 
\end{equation}
that is that of the spin-1/2 Heisenberg antiferromagnet with nearest-neighbor
interactions only and $J_{ij}>0$, it is well known that the zero-temperature 
(at $B=0$) solutions of Hamiltonian (\ref{hh2}) involve radically different 
forms as a function of the geometry, dimensionality, and size.  

Generalizing this behavior to finite magnetic fields $B$, we have found that the
rich variety of the EXD energy spectra presented in Figs.\ \ref{spd30e05} and 
\ref{spd60e05}, as well the EXD spin functions of Section \ref{cpdsb0} can be 
related to those of a 4-site Heisenberg Hamiltonian 
${\cal H}_H^{\text{RP}}(B)$ with $B$-dependent exchange constants 
$\tilde{J}_{ij}(B)=J_{ij}F_{ij}(B)$, and with the four electrons being located 
at the vertices of a rectangular parallelogram (RP) as discussed 
earlier. Due to the reflection symmetry, ${\cal H}_H^{\text{RP}}(B)$ has only 
two different exchange constants $\tilde{J}_{12}=\tilde{J}_{34}$ and 
$\tilde{J}_{14}=\tilde{J}_{23}$, i.e., 
\begin{eqnarray}
{\cal H}_H^{\text{RP}} (B)&=& \tilde{J}_{12}(B) ({\bf S}_1 {\bf \cdot S}_2 +
{\bf S}_3 {\bf \cdot S}_4) + \nonumber \\
&& \tilde{J}_{14}(B) ({\bf S}_1 {\bf \cdot S}_4 + {\bf S}_2 {\bf \cdot S}_3),
\label{hh3}
\end{eqnarray}
where $1 \rightarrow I$, $2 \rightarrow II$, $3 \rightarrow III$, and 
$4 \rightarrow IV$ [in a clockwise direction, see Fig.\ \ref{s0b0}(a)]. Since the
$B=0$ spin exchange interaction constants $J_{ij}$ are expected to decrease 
exponentially\cite{matv09} with the distance between the two sites 
$i$ and $j$, one expects that the Heisenberg model will reproduce the present 
EXD results in the regime $J_{12} << J_{14}$.

To proceed, it is sufficient to use the six-dimensional Ising Hilbert 
subspace for zero total-spin projection ($S_z=0$), which is spanned by the 
following set of basis states [we follow here the ordering in Eq.\ (\ref{x00})]: 
$|1\rangle \rightarrow | \uparrow\uparrow\downarrow\downarrow\rangle$,
$|2\rangle \rightarrow | \uparrow\downarrow\uparrow\downarrow\rangle$, 
$|3\rangle \rightarrow | \uparrow\downarrow\downarrow\uparrow\rangle$,
$|4\rangle \rightarrow | \downarrow\uparrow\uparrow\downarrow\rangle$,
$|5\rangle \rightarrow | \downarrow\uparrow\downarrow\uparrow\rangle$, and
$|6\rangle \rightarrow | \downarrow\downarrow\uparrow\uparrow\rangle$. In
this subspace, the Heisenberg Hamiltonian given by Eq. (\ref{hh3})] can be 
written in matrix form as
\begin{widetext}
\begin{equation} 
{\cal H}_H^{\text{RP}} (B)= \frac{1}{2} \left(
\begin{array}{cccccc} 
\tilde{J}_{12} - \tilde{J}_{14} & \tilde{J}_{14} & 0 & 
0 & \tilde{J}_{14} & 0  \\
\tilde{J}_{14} & - (\tilde{J}_{12} + \tilde{J}_{14}) & \tilde{J}_{12} & 
\tilde{J}_{12} & 0 & \tilde{J}_{14} \\
0 & \tilde{J}_{12} & \tilde{J}_{14} - \tilde{J}_{12} & 
0 & \tilde{J}_{12} & 0  \\
0 & \tilde{J}_{12} & 0 &  
\tilde{J}_{14} - \tilde{J}_{12} & \tilde{J}_{12} & 0 \\
\tilde{J}_{14} & 0 & \tilde{J}_{12} &
\tilde{J}_{12} & - (\tilde{J}_{12} + \tilde{J}_{14}) & \tilde{J}_{14} \\
0 & \tilde{J}_{14} & 0 &
0 & \tilde{J}_{14} &  \tilde{J}_{12} - \tilde{J}_{14} 
\end{array} 
\right).
\label{hhmat} 
\end{equation} 
\end{widetext}

A lengthy, but straightforward, calculation yields the general eigenvalues 
${\cal E}_i$ and corresponding eigenvectors ${\cal V}_i$ of the matrix 
(\ref{hhmat}). The eigenvalues are:
\begin{equation}
{\cal E}_1=-(\tilde{J}_{14} + \tilde{J}_{12})/2,
\label{e1}
\end{equation}
\begin{equation}
{\cal E}_2=(\tilde{J}_{14} - \tilde{J}_{12})/2,
\label{e2}
\end{equation}
\begin{equation}
{\cal E}_3=(\tilde{J}_{12} - \tilde{J}_{14})/2,
\label{e3}
\end{equation}
\begin{equation}
{\cal E}_4=(\tilde{J}_{14} + \tilde{J}_{12})/2,
\label{e4}
\end{equation}
\begin{equation}
{\cal E}_5=-(\tilde{J}_{14} + \tilde{J}_{12})/2
-Q(\tilde{J}_{14},\tilde{J}_{12}),
\label{e5}
\end{equation}
\begin{equation}
{\cal E}_6=-(\tilde{J}_{14} + \tilde{J}_{12})/2
+Q(\tilde{J}_{14},\tilde{J}_{12}),
\label{e6}
\end{equation}
where 
\begin{equation}
Q(a,b)=\sqrt{a^2-ab+b^2}.  
\label{qq}
\end{equation}

The corresponding unnormalized eigenvectors and their total spins are given by:
\begin{equation}
{\cal V}_1=
\{0, -1, 0, 0, 1, 0 \},\;\;S=1,
\label{v1}
\end{equation}
\begin{equation}
{\cal V}_2=
\{0, 0, -1, 1, 0, 0 \},\;\;S=1,
\label{v2}
\end{equation}
\begin{equation}
{\cal V}_3=
\{ -1, 0, 0, 0, 0, 1 \},\;\;S=1,
\label{v3}
\end{equation}
\begin{equation}
{\cal V}_4=
\{1, 1, 1, 1, 1, 1 \},\;\;S=2,
\label{v4}
\end{equation}
\begin{equation}
{\cal V}_5=
\{1, -{\cal X}, -1+{\cal X}, -1+{\cal X}, -{\cal X}, 1 \},\;\; S=0,
\label{v5}
\end{equation}
\begin{equation}
{\cal V}_6=
\{1, -{\cal Y}, -1+{\cal Y}, -1+{\cal Y}, -{\cal Y}, 1 \},\;\; S=0,
\label{v6}
\end{equation}
where 
\begin{equation}
{\cal X}=r+Q(1,r),
\label{xq}
\end{equation}
\begin{equation}
{\cal Y}=r-Q(1,r),
\label{yq}
\end{equation}
and $r=\tilde{J}_{12}/\tilde{J}_{14}$.

To understand how the Heisenberg Hamiltonian in Eq.\ (\ref{hh3}) captures
the rich behavior seen in the EXD spectra of Figs. \ref{spd30e05} and 
\ref{spd60e05}, we start with the limiting case $r \rightarrow 0$, which is 
applicable (see below) to the larger interdot distance $d=60$ nm. In this limit,
one can neglect $\tilde{J}_{12}$ compared with $\tilde{J}_{14}$, which results 
in partial degeneracies within the band; namely one has
${\cal E}_2={\cal E}_4={\cal E}_6=\tilde{J}_{14}/2$,
${\cal E}_1={\cal E}_3=-\tilde{J}_{14}/2$, and
${\cal E}_5=-3\tilde{J}_{14}/2$.
This $3-2-1$ degeneracy pattern is independent of the magnetic-field
dependence through $F_{14}(B)$ [$\tilde{J}_{ij}=J_{ij} F_{ij}(B)$]
and is characteristic of all three EXD spectra
(for $\kappa=12.5$, 6, and 2) associated with the larger interdot distance
$d=60$ nm. Furthermore, the fact that all six curves in the EXD lowest-energy 
band appear to cross at the same point $B^C(\kappa)$ (reversing at the same 
time the order of the degenerate levels) suggests that
\begin{equation} 
F_{14}(B) \sim \cos[\pi B/(2 B_{14}^C)]. 
\label{f14}
\end{equation}
It is remarkable that the behavior described above is prominent even for the 
weak interelectron repulsion $\kappa=12.5$ [see Fig.\ \ref{spd60e05}(a)] when 
the extent of electron localization and the formation of a Wigner molecule are 
not clearly visible via an inspection of the corresponding electron densities 
[see Fig.\ \ref{elden30_60}(b)].

Of interest is the fact that the ability of the Heisenberg Hamiltonian
in Eq.\ (\ref{hh3}) to reproduce the EXD trends is not restricted solely to
energy spectra, but extends to the EXD wave functions as well. Indeed when
$\tilde{J}_{12} \rightarrow 0$, the last two eigenvectors of the Heisenberg 
matrix (having $S=0$) become
\begin{equation}
{\cal V}_5 \rightarrow \{1,-1,0,0,-1,1\},
\label{v52}
\end{equation}
and
\begin{equation}
{\cal V}_6 \rightarrow \{1,1,-2,-2,1,1\}.
\label{v62}
\end{equation}
When multiplied by the normalization factor, the wave functions represented by 
the eigenvectors in Eqs.\ (\ref{v52}) and (\ref{v62}) coincide (within an 
overall $\mp 1$ sign) with the EXD spin functions ${\cal X}_{00}^{(1)}$ and 
${\cal X}_{00}^{(2)}$ in Eqs.\ (\ref{x00_cpd60k2b0l1}) and 
(\ref{x00_cpd60k2b0l3}), respectively. In addition, when again multiplied
by the corresponding normalization factor, the wave function represented by the
eigenvector ${\cal V}_1$ [Eq.\ (\ref{v1})] (having total spin $S=1$) coincides
(within an overall minus sign) with the EXD spin function ${\cal X}_{10}$ in
Eq.\ (\ref{x10_cpd60k2b0l2}).

The EXD spectra and spin functions for the shorter distance $d=30$ nm can
be analyzed within the framework of the 4-site Heisenberg Hamiltonian
(\ref{hhmat}) when small (compared with $\tilde{J}_{14}$), but nonnegligible, 
values of the second exchange integral $\tilde{J}_{12}$ are considered. In this 
case, the partial three-fold and two-fold degeneracies are lifted. Indeed in 
Figs.\ \ref{spd30e05}(a) ($\kappa=12.5$) and \ref{spd30e05}(b) ($\kappa=6$), the
EXD lowest-energy band consists of six distinct levels. For the strong 
interelectron case with $\kappa=2$ and $d=30$ nm [Fig.\ \ref{spd30e05}(c)], 
however, the EXD spectra indicate that the effective (selfconsistent-field) 
potential barrier between the dots due to the Coulomb repulsion is high enough 
to reduce $\tilde{J}_{12}$ to a negligeable value and to produce spectra 
exhibiting the characteristic $3-2-1$ degeneracy pattern that is prominent in 
the spectra associated with the larger interdot distance $d=60$ nm. (For the 
wave functions in the $\kappa=2$ and $d=30$ nm case, however, the influence of 
$\tilde{J}_{12}$ cannot be neglected, see below.) 

In addition to the lifting of the partial degeneracies within the lowest-energy
band, one observes from an examination of Figs.\ \ref{spd30e05}(a) and 
\ref{spd30e05}(b) the occurrence of two characteristic secondary 
oscillations (as a function of $B$) emerging out of the previously three-fold 
and two-fold degenerate levels. Using the Heisenberg Hamiltonian, these 
secondary oscillations can be described by taking the second exchange constant
to have a $B$-dependence similar to that in Eq.\ (\ref{f14}), i.e.,  
\begin{equation} 
F_{12}(B) \sim \cos[\pi B/(2 B_{12}^C)+\phi_0], 
\label{f12}
\end{equation}
with $B_{12}^C < B_{14}^C$, $\phi_0$ being a phase shift.
This secondary oscillation is superimposed on the
main oscillation specified by $F_{14}(B)$ [Eq.\ (\ref{f14})] in
accordance with the expressions for the Heisenberg energy levels
${\cal E}_2$ [Eq.\ (\ref{e2})], ${\cal E}_4$ [Eq.\ (\ref{e4})], 
${\cal E}_6$ [Eq.\ (\ref{e6})], and ${\cal E}_1$ [Eq.\ (\ref{e1})],
${\cal E}_3$ [Eq.\ (\ref{e3})].

Another characteristic feature developing for $d=30$ nm in Figs.\ 
\ref{spd30e05}(a) and \ref{spd30e05}(b)
is the anticrossing gap $\Delta$ between the two $S=0$ states. According
to the Heisenberg model this gap is given by 
$\Delta=2 J_{12} \cos[\pi B_{14}^C/(2B_{12}^C)+\phi_0]$.
As a concrete example of the above, we estimated that the spectrum of Fig.\
\ref{spd30e05}(b) can be rather well reproduced using the expressions for the 
Heisenberg eigenvalues when $J_{12}/J_{14} \approx -1/4.1$, $B_{12}^C 
\approx B_{14}^C/2.5$, and $\phi_0=\pi/2.4$.    

Similar to the findings in the larger-distance ($d=60$ nm) case, the agreement
between EXD and Heisenberg-model results in the smaller $d=30$ nm distance
includes also the wave functions. Indeed, as discussed in Section 
\ref{sec_s_one}, the $P_{xy}=1$, $S=1$ EXD spin function for $d=30$ nm 
(and $B=0$, $\kappa=2$) was found to be identical to the one determined
for the larger interdot distance $d=60$ nm,
i.e., it is given by expression (\ref{x10_cpd60k2b0l2}). This reflects
the remarkable property that the Heisenberg eigenvector ${\cal V}_1$ 
[Eq.\ (\ref{v1})] is independent of the two exchange constants 
$\tilde{J}_{14}$ and $\tilde{J}_{12}$, and thus independent of the
interdot distance $d$ (as well as of the dielectric constant $\kappa$ and the
magnetic field $B$). On the other hand, the $S=0$ Heisenberg eigenvectors  
[Eqs.\ (\ref{v5}) and (\ref{v6})] do depend on the ratio $r=\tilde{J}_{12}/
\tilde{J}_{14}$, which is in agreement with the fact that the EXD 
ground-state spin function ${\cal X}_{00}^{(3)}$ for $d=30$ nm in Eq.\ 
(\ref{x00_cpd30k2b0l1}) is slightly different from the corresponding
$S=0$ EXD spin function ${\cal X}_{00}^{(1)}$ for $d=60$ nm [see Eq.\ 
(\ref{x00_cpd60k2b0l1})]. With consideration of the normalization factor,
we estimate that the Heisenberg eigenvector ${\cal V}_5$ [Eq.\ (\ref{v5})] 
agrees with ${\cal X}_{00}^{(3)}$ when $r \approx -1/7.5$.  

It is of interest to contrast the EXD spin functions determined in Section 
\ref{cpdsb0} with the well known solutions of the Heisenberg Hamiltonian
[Eq.\ (\ref{hh3})] when the four spins are located on four sites arranged in a
perfect square,\cite{haas07,fazebook} i.e., when the two exchange integrals
are equal; $\tilde{J}_{14}=\tilde{J}_{12}=J>0$. (The perfect-square arrangement 
arises \cite{shi07} also in the case of formation of a four-electron Wigner 
molecule in a single circular quantum dot.) In this case, the ground state of 
${\cal H}_H$ is the celebrated resonating valence bond (RVB) 
state\cite{haas07,fazebook} which forms the basic unit block in many theoretical 
approaches aiming at describing high-temperature superconductors.\cite{ande08} 
The RVB state has quantum numbers $S=0$, $S_z=0$ and a Heisenberg energy
${\cal E}_5=-2J$; it is given by the normalized version of ${\cal V}_5$ 
[Eq.\ (\ref{v5})] when $r=1$, that is by (see also Refs.\
\onlinecite{haas07} and \onlinecite{fazebook})
\begin{eqnarray} 
{\cal X}_{00}^{\text{RVB}} &=&
\frac{1}{2\sqrt{3}} |\uparrow\uparrow\downarrow\downarrow\rangle
-\frac{1}{\sqrt{3}} |\uparrow\downarrow\uparrow\downarrow\rangle
+\frac{1}{2\sqrt{3}} |\uparrow\downarrow\downarrow\uparrow\rangle \nonumber \\
&& +\frac{1}{2\sqrt{3}} |\downarrow\uparrow\uparrow\downarrow\rangle
-\frac{1}{\sqrt{3}} |\downarrow\uparrow\downarrow\uparrow\rangle
+\frac{1}{2\sqrt{3}} |\downarrow\downarrow\uparrow\uparrow\rangle.\nonumber \\
\label{x00_rvb}
\end{eqnarray}

Although the (excited-state) EXD ${\cal X}_{00}^{(2)}$ [Eq.\ 
(\ref{x00_cpd60k2b0l3})] in the quantum-double-dot case portrayed in Fig.\ 
\ref{s0b0}(c) appears (superficially) to be similar to the
(ground-state) RVB ${\cal X}_{00}^{\text{RVB}}$ [Eq.\ (\ref{x00_rvb})], the two
are not equal. Indeed the coefficients of the pair of Slater determimants
$|\uparrow\downarrow\uparrow\downarrow\rangle$ and 
$|\downarrow\uparrow\downarrow\uparrow\rangle$ have been interchanged with those
of $|\downarrow\uparrow\uparrow\downarrow\rangle$ and 
$|\uparrow\downarrow\downarrow\uparrow\rangle$. 
Similar observations apply also to the remaining pair of $S=0$ and $S_z=0$ 
states that are orthogonal to ${\cal X}_{00}^{(2)}$ [see ${\cal X}_{00}^{(1)}$ in
Eq.\ (\ref{x00_cpd60k2b0l1}); case of double quantum dot] and to   
${\cal X}_{00}^{\text{RVB}}$ (case of a perfect square), respectively. 



We note that the differences in the ${\cal X}_{00}$ spin functions between the 
DQD case (corresponding to a rectangular parallelogram) and the perfect-square
case are also reflected in the ${\cal P}_{\uparrow\downarrow}$ CPDs. Indeed the 
CPDs of the DQD (Fig.\ \ref{s0b0}) exhibit equal-height humps along the 
smaller side of the parallelogram, while those 
of the perfect-square configuration (and/or circular
quantum dot) exhibit equal-height humps along a diagonal.\cite{shi07}

\section{Discussion}

\subsection{Magnetic-field dependence and relevance to quantum computing}
\label{disca}

Strongly correlated electrons on a lattice are frequently described by the 
Hubbard-model Hamiltonian
\begin{equation}
H^{\text{Hubbard}}=
-\sum_{i,j,\sigma} t_{ij} c^\dagger_{i\sigma} c_{j\sigma}
+U \sum_i n_{i\uparrow} n_{i\downarrow},
\label{hhubb}
\end{equation}
where $t_{ij}$ is the hopping integral from site $j$ to site $i$, $\sigma=\pm1$ 
[equivalently this denotes a spin up ($\uparrow$) or a spin down 
($\downarrow$)], and $n_{i\sigma}=c^\dagger_{i\sigma} c_{i\sigma}$,
with $c^\dagger_{i\sigma}$ and $c_{i\sigma}$ being single-particle 
creation and annihilation operators for the site $i$. $U$ is the on-site
Coulomb repulsion.

It is well known that the one-band Hubbard model at half-filling reduces 
\cite{emer76,vief04} (to lowest order) to a Heisenberg antiferromagnetic 
Hamiltonian ${\cal H}_H$ [see Eq.\ (\ref{hh2})] in the limit of the on-site 
Coulomb repulsion $U>0$ being large relative to the hopping integral $t_{ij}$. 
In the absence of an applied magnetic field ($B=0$), $t_{ij}$ can be taken to be
real and the corresponding exchange integrals are given by \cite{emer76} 
\begin{equation}
J_{ij}^{\text{Hubbard}}=2t_{ij}^2/U,
\label{jhubb}
\end{equation}
 
In the presence of a magnetic field (which is the case of this paper), $t_{ij}$ 
picks up \cite{sen95,rokh9091} a Peierls phase 
$\exp[(ie/c\hbar) \int_i^j {\bf A} \cdot d{\bf r}]$, where ${\bf A}$ is the 
vector potential. In this case $t_{ij}$ is complex, and one must replace 
\cite{sen95,rokh9091} $t_{ij}^2 \rightarrow t_{ij}t_{ij}^*$ in Eq.\ 
(\ref{jhubb}). The complex conjugation, however, cancels any magnetic-field 
effect associated with the Peierls phase factor, which means that the 
$J_{ij}^{\text{Hubbard}}$ are independent of $B$.

In sharp contrast with this Hubbard-model result, our EXD calculations indicate
that the exchange integrals entering in the Heisenberg Hamiltonian 
of Eq.\ (\ref{hh3}) depend strongly on the magnetic field. Such strong 
$B$-dependence of the exchange integrals has been found in previous theoretical
studies in the simpler case of two electrons in double quantum dots,
\cite{loss99, hell05,lebu06,yann02,yann02.2} as well as in anisotropic single 
quantum dots.\cite{yann06,yann07.2} (For two electrons, the exchange integral is
calculated as the energy difference between the singlet and triplet states.) 
This $B$-dependence in the case of two electrons in quantum dots has also been 
observed experimentally.\cite{yann06,yann07.2,marc04} Following an earlier 
proposal,\cite{loss99} the $B$-dependence of $J$ for the two-electron case 
has developed into a central theme in experimental efforts focussing on 
solid-state implementation of quantum computing.\cite{hans07,taru08}
In this context, our EXD results in this paper extend the $B$-dependence
of the exchange integrals to larger numbers $(N > 2)$ of electrons in quantum dot
molecules.

The physics underlying the emergence of such strong $B$-dependence in the case of
solid-state artificial nanostructures is clearly related to the importance 
\cite{kouw01,yann07} of orbital magnetic effects resulting from the much larger 
size (by a factor of 10000) of the electronic wave functions in 2D quantum dots 
compared to that in natural atoms. For spin interactions between electrons 
localized within the natural atoms, huge magnetic fields (of order 
10000 T) are required for reproducing a $B$-dependence of the exchange 
integrals similar to that discussed in this paper.   

\subsection{Aspects of spin entanglement}
\label{discb}

N\'{e}el antiferromagnetic ordering, where the average spin per site 
$<S^z_j>=(-1)^{j+1}/2$,  is an important magnetic phenomenon in the 
thermodymanic limit\cite{fazebook} associated with breaking of the total-spin
symmetry. The finite size magnetic clusters discussed here exhibit a sharply 
different behavior in this respect. Indeed, as
discussed in Ref.\ \onlinecite{fazebook}, the four-site N\'{e}el state is the 
single Slater determinant $|\downarrow\uparrow\downarrow\uparrow\rangle$ (or
$\uparrow\downarrow\uparrow\downarrow\rangle$). It is clear that the total-spin
conserving EXD functions ${\cal X}_{00}$ (as well as the corresponding 
Heisenberg eigenvectors) are multideterminental and 
have an average spin per localized electron (per site) $<S^z_j>=0$.

We concur with Ref.\ \onlinecite{fazebook} that the phenomenon of N\'{e}el 
antiferomagnetism is radically  modified in assemblies of few electrons. In this
section, we argue that instead of ``antiferromagnetic ordering'' the appropriate
physical concept for the WM states found earlier is that of {\it spin 
entanglement\/}. Indeed, in the previous sections, we showed that the EXD wave 
functions in the regime of Wigner-molecule formation can be approximated as a 
superposition of a small number of Slater determinants corresponding to well 
structured spin functions; see, e.g., ${\cal X}_{00}^{(1)}$ in Eq.\ 
(\ref{x00_cpd60k2b0l1}). This is a great
simplification compared to the initial EXD superposition [Eq.\ (\ref{mbwf})],
where the counting index is usually $I \geq 500,000$. This reduction of the
molecular EXD solutions to their equivalent spin functions (described in
Section \ref{cpdsb0}) (or to the Heisenberg eigenvectors described in
Section \ref{heis}) enables one to investigate their properties 
regarding fundamental quantum behavior associated with quantum
correlations and fluctuations beyond the mean field.

The mathematical theory of entanglement is still developing and includes 
several directions. One way to study entanglement is through the use of properly
defined measures of entanglement, e.g., the von Neumann entropy which utilizes 
the single-particle density matrix. Another way is to catalog and specify  
classes of entangled states that share common properties regarding 
multipartite entanglement. A well known class of $N$-qubit entangled states are 
the Dicke states,\cite{dick54,vers02,stoc03,korb06} which most often are
taken to have the symmetric form:
\begin{equation}
{\cal X}^{\text{Dicke}}_{N,k} =
\left( \begin{array}{c} N \\ k \end{array} \right)^{-1/2}
( |\underbrace{11\ldots1}_k00\ldots0\rangle + {\text{Perm}} ).
\label{dicke_sym}
\end{equation}
Each qubit is a linear superposition of two single-particle states denoted by
0 or 1, and the symbol 'Perm' stands for all remaining permutations. The
0 or 1 do not have to be necessarily up or down 1/2-spin states. Two-level
atoms in linear ultracold traps have already been used as an implementation of a 
qubit. Dicke states appear in many physical processes like superradiance
and superfluorescence. They can also be realized with photons, where the qubits
correspond to the polarization degree of freedom.\cite{korb06}

In the 1/2-spin case of fermions (e.g., for electrons), the Dicke states
of Eq.\ (\ref{dicke_sym}) correspond to a fully symmetric flip of $k$
out of $N$ localized spins. It is apparent that the four-qubit fully polarized
($S=2$ with spin projection $S_z$=0) EXD solution is reproduced by 
${\cal X}_{20}$ of Eq.\ (\ref{x20}), 
and thus it is of the symmetric Dicke form (with $k=2$) 
displayed above in Eq.\ (\ref{dicke_sym}). On the other part, the DQD EXD states
(with $S_z=0$) studied in Section \ref{cpdsb0} with $S=0$ and/or $S=1$ 
represent a natural generalization of Eq.\ (\ref{dicke_sym}) to the class of 
{\it asymmetric\/} Dicke states. 

Dicke states with a single flip ($k=1$) are known as $W$ 
states.\cite{cira00,woot00} For $N=4$ electrons, the 
latter states are related to EXD solutions with $S_z=\pm 1$. For the connection 
between $W$ states and EXD states for $N=3$ electrons in anisotropic quantum 
dots, see Ref.\ \onlinecite{yues07}. $W$ states have already been realized 
experimentally using two-level ultracold ions in linear traps.\cite{roos05}

\section{Summary}

Extensive investigations of lateral double quantum dots containing four 
electrons ({\it artificial quantum-dot Helium molecules\/}) were performed using
the exact-diagonalization method (described in Section \ref{sec:3}), as a 
function of interdot separation, applied magnetic field, and strength of 
interelectron repulsion. Novel quantum behavior was discovered compared to 
circular QDs concerning energy spectra, analogies with finite Heisenberg 
clusters, and aspects of entanglement. It is hoped that the present work will 
motivate further experimental studies on lateral DQDs with more than two 
electrons.

Specifically it was found (Section \ref{exdspec}) that, 
as a function of the magnetic field, the energy 
spectra exhibit a low-energy band consisting of a group of six states, and that  
this number six is not accidental, but a consequence of the conservation of the 
total spin and of the ensuing spin degeneracies and supermultiplicities 
expressed in the branching diagram (described in Section \ref{mbsfs}). 
These six states appear to cross at a single
value of the magnetic field, and the crossing point gets sharper for larger 
interdot distances. As the strength of the Coulomb repulsion increases, the six 
states tend to become degenerate and a well defined energy gap separates them 
from the higher-in-energy excited states. 

The formation of the low-energy band is a consequence of the localization of
the four electrons within each dot (with two electrons on each dot). The 
result is formation (with increasing strength of the Coulomb  repulsion) of a 
Wigner supermolecule,  with the four localized electrons at the 
corners of a rectangular parallelogram. Using the spin-resolved pair-correlation
functions, it was shown that one can map the EXD many-body wave functions to the
spin functions associated with four localized spins (Section \ref{cpdsb0}). 

This mapping led us naturally to studying analogies with finite systems 
described by model Heisenberg Hamiltonians (referred to often as finite 
Heisenberg clusters). Specifically, we provided a detailed interpretation of the
EXD spin functions and EXD spectra associated with the low-energy band via a 
4-site finite Heisenberg cluster characterized by two (intradot and interdot) 
exchange integrals. More importantly, our EXD calculations suggest a prominent 
oscillatory magnetic-field dependence of the two exchange integrals entering in 
this 4-site Heisenberg Hamiltonian (Section \ref{heis}).

Such strong $B$-dependence of the exchange 
integrals has been found in previous theoretical and experimental studies in the
simpler case of two electrons in quantum dots, and it has developed into a 
central theme in experimental efforts aiming at solid-state implementation of 
quantum computing. Our EXD results in this paper extend the $B$-dependence of 
the exchange integrals to larger numbers $(N > 2)$ of electrons in quantum dot 
molecules (see discussion in Section \ref{disca}).

Finally, it was discussed that the EXD spin functions correspond to strongly 
entangled states known in the literature of quantum information as $N$-qubit 
Dicke states (Section \ref{discb}).

\begin{acknowledgments}
This work was supported by the US D.O.E. (Grant No. FG05-86ER45234).
\end{acknowledgments}

\appendix*

\section{Single-particle states of the two-center oscillator}

\begin{figure}[t]
\centering{
\includegraphics[width=8.0cm]{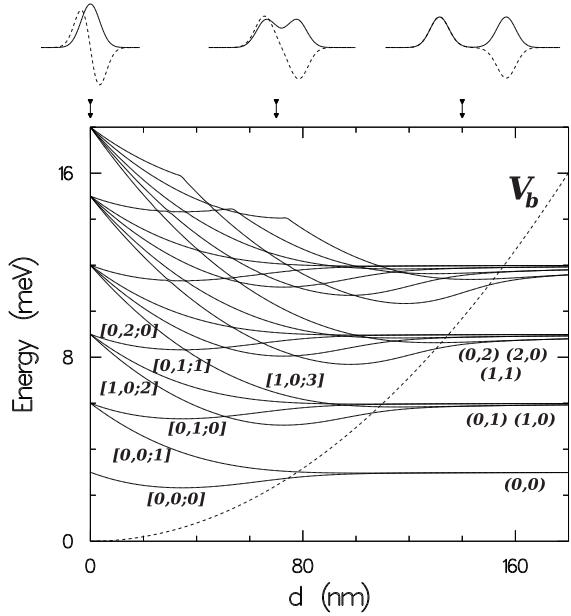}} 
\caption{Single-particle spectra of a double quantum at $B=0$ plotted versus the
distance $d$ between two (identical) coupled QDs with a TCO confinement 
$\hbar \omega_{x1}= \hbar \omega_{x2}=\hbar \omega_y =3$ meV and $h_1=h_2=0$
[see Eq.\ (\ref{hsp})]. For all $d$'s the barrier control
parameters were taken as $\epsilon_1^b=\epsilon_2^b=0.5$, i.e., the barrier
height (depicted by the dashed line) varies as $V_b(d)=V_0(d)/2$.
Molecular orbitals correlating the united ($V_b=0$) and separated-dots limits
are denoted along with the corresponding (on the right) single-QD states.
Wave function cuts at $y=0$ along the $x$-axis at several distances $d$
(see arrows) corresponding to the lowest bonding and antibonding 
eigenvalues (solid and dashed lines, respectively) are displayed at the top. 
Energies in meV and distances in nm.
}
\label{espd}
\end{figure}

\begin{figure}[t]
\centering{
\includegraphics[width=8.0cm]{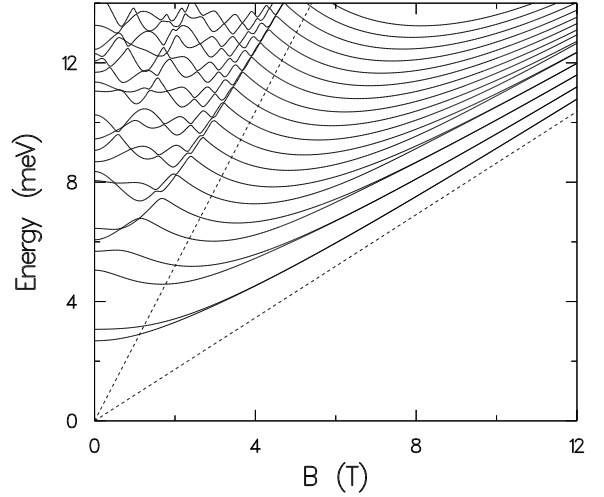}} 
\caption{Single-particle spectrum of the $d=70$ nm ($\hbar \omega_{x1}=
\hbar \omega_{x2}=$ $\hbar \omega_y=3$ meV, $V_b=2.43$ meV, $h_1=h_2=0$) double 
quantum dot versus $B$ (in T). The $\hbar \omega_c/2$ (${\cal N}_L=0$, lower
line) and $3\hbar \omega_c/2$ (${\cal N}_L=1$, upper line) first and second 
Landau levels are given by the dashed lines. 
}
\label{espb}
\end{figure}

In this Appendix, we discuss briefly the energy spectra associated with 
the single-particle states of the two-center oscillator Hamiltonian 
given by Eq.\ (\ref{hsp}). We follow here the notation presented first in
Ref.\ \onlinecite{yl99}. For further details, see Ref.\ \onlinecite{note2}. 

The calculated two-center oscillator 
single-particle spectrum for a double quantum dot made of two tunnel-coupled 
identical QDs (with $\hbar \omega_y = \hbar \omega_{x1} = \hbar \omega_{x2}=3$ 
meV) plotted versus the distance, $d$, between the centers of the two dots, 
is given in Fig.\ \ref{espd}.  In these calculations, the height of the barrier 
between the dots varies as a function of $d$, thus simulating reduced tunnel
coupling between them as they are separated; we take the barrier control 
parameter as $\epsilon_1^b = \epsilon_2^b = 0.5$. In the calculations in this
Appendix, we used GaAs values, $m^*=0.067 m_e$ and a dielectric constant 
$\kappa=12.9$. For the separated single QDs (large $d$) and the unified QD 
($d = 0$) limits, the spectra are the same, corresponding to that of a 2D 
harmonic oscillator (being doubly degenerate for the separated single QDs)
with a level degeneracy of 1, 2, 3, ... .  
In analogy with real molecules, the single-particle states in the intermediate 
region ($d > 0$) may be interpreted as molecular orbitals (MOs) 
made of linear superpositions of the states of the two dots comprising the 
DQ  This qualitative description is intuitively appealing, though it is more 
appropriate for the weaker coupling regime (large $d$); nevertheless we continue
to use it for the whole range of tunnel-coupling strengths between the dots, 
including the strong coupling regime where reference to the states of the 
individual dots is only approximate.  Thus, for example, as the two dots 
approach each other, the lowest levels ($n_x$, $n_y$) with $n_x = n_y = 0$ on 
the two dots may combine symmetrically (``bonding'') or antisymmetrically 
(``antibonding'') to form [0,0;0] and [0,0;1] MOs, with the third index denoting
the total number of nodes of the MO along the interdot axis ($x$), that is, 
$2 n_x +{\cal I}$, ${\cal I}=$ 0 or 1; for symmetric combinations 
(${\cal I}=0$), this index is even and for antisymmetric ones (${\cal I}=1$), it
is odd.  Between the separated-single-QDs and the unified-QD limits, the 
degeneracies of the individual dots' states are lifted, and in correlating 
these two limits the number of $x$-nodes is conserved; for example the [0,0;1] 
MO converts in the unified-QD limit into the (1,0) state of a single QD, the 
[1,0;2] MO into the (2,0) state, and the [0,1;1] MO 
into the (1,1) state (see Fig.\ \ref{espd}). Note that MOs of different 
symmetries may cross, while they do not if they are of the same symmetry.  

In a magnetic field, the TCO model consitutes a generalization of the 
Darwin-Fock model \cite{fd} for non-interacting electrons in a single circular
QD. The single-particle spectra for the DQD ($d = 70$ nm, $V_b = 2.43$
meV) in a magnetic field ($B$) are shown in Fig.\ \ref{espb}
(here we neglect the Zeeman interaction which is small for our range of $B$
values with $g^*=-0.44$ for GaAs).  
The main features are: (i) the multiple crossings (and avoided crossings) as 
$B$ increases, (ii) the decrease of the energy gap between levels, occurring in 
pairs (such as the lowest bonding-antibonding pair), portraying an effective 
reduced tunnel coupling between the QDs comprising the DQD as $B$ increases, 
(iii) the ``condensation'' of the spectrum into the sequence of Landau levels 
$({\cal N}_L + 1/2)\hbar \omega_c$, ${\cal N}_L=$ 0, 1, 2,
... (the ${\cal N}_L=0$ and ${\cal N}_L=1$ bands are depicted, respectively,
by the lower and upper dashed lines in Fig.\ \ref{espb}). This is 
similar to the behavior of the single-particle Darwin-Fock spectrum for 
harmonically confined electrons in a circular QD \cite{fd}
(note however that the geometry of the DQD is non-circular and deviates from a 
simple harmonic confinement).


\begin{thebibliography}{999}
\bibitem{note1}
A three-dimensional magnetic-field-free version of the TCO has been
used in the description of fission in metal clusters \cite{yl95}
and atomic nuclei. \cite{grei72}
\bibitem{note2}
The two-dimensional TCO was used earlier to approximate the solutions of 
the many-body Hamiltonian in Eq.\ (\ref{mbh}) at the level of unrestricted 
Hartree-Fock (UHF) exhibiting breaking of total-spin symmetry, see
Ref.\ \onlinecite{yann99}, Ref.\ \onlinecite{yann02}, and
Ref.\ \onlinecite{yann02.2}.
For a review of the necessary subsequent step of symmetry
restoration, see Ref.\ \onlinecite{yann07}.
\bibitem{note5}
Due to the very long CPU times required for calculating surfaces, several
CPDs at $B=0$ were calculated using $K=32$. We checked that this does not
influence our analysis of spin functions compared to $K=50$.
\end{thebibliography}
\end{document}